\documentclass[twocolumn,times]{aastex7}
\usepackage{graphicx}
\usepackage{natbib}
\usepackage{color}
\usepackage{wrapfig}
\usepackage{xspace}
\usepackage{multirow}
\usepackage{amsmath}

\graphicspath{{./}{figures/}}

\newcommand{\mj}{\ensuremath{\,M_{\rm Jup}}}

\newcommand{\mearth}{$M_\oplus$}
\newcommand{\rearth}{$R_\oplus$}
\newcommand{\acen}{$\alpha$~Cen\xspace}
\newcommand{\acenA}{$\alpha$~Cen~A\xspace}
\newcommand{\emus}{$\epsilon$~Mus\xspace}
\newcommand{\acenB}{$\alpha$~Cen~B\xspace}
\newcommand{\acenAB}{$\alpha$~Cen~AB\xspace}

\newcommand{\webbpsf}{\texttt{STPSF}\xspace}
\newcommand{\mum}{$\mu$m}

\newcommand{\objS}{\emph{S}1\xspace}

\newcommand{\objC}{\emph{C}1\xspace}

\begin{document}

\title{Worlds Next Door: A Candidate Giant Planet Imaged in the Habitable Zone of \acenA. \\ II. Binary Star Modeling, Planet and Exozodi Search, and Sensitivity Analysis}

\correspondingauthor{Aniket Sanghi \& Charles Beichman}

\author[0000-0002-1838-4757]{Aniket Sanghi}
\altaffiliation{Shared first authorship.}
\affiliation{Cahill Center for Astronomy and Astrophysics, California Institute of Technology, 1200 E. California Boulevard, MC 249-17, Pasadena, CA 91125, USA}
\affiliation{NSF Graduate Research Fellow}
\email[show]{asanghi@caltech.edu}

\author[0000-0002-5627-5471]{Charles Beichman}
\altaffiliation{Shared first authorship.}
\affiliation{NASA Exoplanet Science Institute, Caltech-IPAC, Pasadena, CA 91125, USA}
\affiliation{Jet Propulsion Laboratory, California Institute of Technology, Pasadena, CA 91109, USA}
\email[show]{chas@ipac.caltech.edu}

\author[0000-0002-8895-4735]{Dimitri Mawet}
\affiliation{Cahill Center for Astronomy and Astrophysics, California Institute of Technology, 1200 E. California Boulevard, MC 249-17, Pasadena, CA 91125, USA}
\affiliation{Jet Propulsion Laboratory, California Institute of Technology, Pasadena, CA 91109, USA}
\email{dmawet@astro.caltech.edu}

\author[0000-0001-6396-8439]{William O. Balmer}
\affiliation{Department of Physics \& Astronomy, Johns Hopkins University, 3400 N. Charles Street, Baltimore, MD 21218, USA}
\affiliation{Space Telescope Science Institute, 3700 San Martin Drive, Baltimore, MD 21218, USA}
\email{wbalmer1@jhu.edu}

\author[0000-0003-0958-2150]{Nicolas Godoy}
\affiliation{Aix Marseille Univ., CNRS, CNES, LAM, Marseille, France}
\email{nicolas.godoy@lam.fr}

\author[0000-0003-3818-408X]{Laurent Pueyo}
\affiliation{Space Telescope Science Institute, 3700 San Martin Drive, Baltimore, MD 21218, USA}
\email{pueyo@stsci.edu}

\author[0000-0001-9353-2724]{Anthony Boccaletti}
\affiliation{LIRA, Observatoire de Paris, Universit\'e PSL, Sorbonne Universit\'e, Universit\'e Paris Cit\'e, CY Cergy Paris Universit\'e, CNRS, 5 place Jules Janssen, 92195 Meudon, France}
\email{anthony.boccaletti@obspm.fr}

\author[0000-0003-4761-5785]{Max Sommer}
\affiliation{Institute of Astronomy, University of Cambridge, Madingley Road, Cambridge CB3 0HA, UK}
\email{ms3078@cam.ac.uk}

\author[0000-0002-9799-2303]{Alexis Bidot}
\affiliation{Space Telescope Science Institute, 3700 San Martin Drive, Baltimore, MD 21218, USA}
\email{abidot@stsci.edu}

\author[0000-0002-9173-0740]{Elodie Choquet}
\affiliation{Aix Marseille Univ., CNRS, CNES, LAM, Marseille, France}
\email{elodie.choquet@lam.fr}

\author[0000-0003-0626-1749]{Pierre Kervella}
\affiliation{LIRA, Observatoire de Paris, Universit\'e PSL, Sorbonne Universit\'e, Universit\'e Paris Cit\'e, CY Cergy Paris Universit\'e, CNRS, 5 place Jules Janssen, 92195 Meudon, France}
\affiliation{French-Chilean Laboratory for Astronomy, IRL 3386, CNRS and U. de Chile, Casilla 36-D, Santiago, Chile}
\email{pierre.kervella@obspm.fr}

\author{Pierre-Olivier Lagage}
\affiliation{Universit\'e Paris-Saclay, Universit\'e Paris Cit\'e, CEA, CNRS, AIM, 91191 Gif-sur-Yvette, France}
\email{pierre-olivier.lagage@cea.fr}

\author[0000-0002-0834-6140]{Jarron Leisenring}
\affiliation{Steward Observatory, University of Arizona, Tucson, AZ 85721, USA}
\email{jarronl@arizona.edu}

\author[0000-0002-3414-784X]{Jorge Llop-Sayson}
\affiliation{Jet Propulsion Laboratory, California Institute of Technology, Pasadena, CA 91109, USA}
\email{jorge.llop.sayson@jpl.nasa.gov}

\author[0000-0001-5644-8830]{Michael Ressler}
\affiliation{Jet Propulsion Laboratory, California Institute of Technology, Pasadena, CA 91109, USA}
\email{michael.e.ressler@jpl.nasa.gov}

\author[0000-0002-4309-6343]{Kevin Wagner}
\affiliation{Steward Observatory, University of Arizona, Tucson, AZ 85721, USA}
\email{kevinwagner@arizona.edu}

\author[0000-0001-9064-5598]{Mark Wyatt}
\affiliation{Institute of Astronomy, University of Cambridge, Madingley Road, Cambridge CB3 0HA, UK}
\email{wyatt@ast.cam.ac.uk}

\shorttitle{JWST/MIRI Observations of \acenA: Paper II}
\shortauthors{Sanghi \& Beichman et al.}

\begin{abstract}
The \emph{James Webb} Space Telescope (JWST) observed our closest solar twin, $\alpha$~Centauri~A (\acenA), with the Mid-InfraRed Instrument (MIRI) in the F1550C (15.5 $\mu$m) coronagraphic imaging mode at three distinct epochs between August 2024 and April 2025. For the first time with JWST, we demonstrate the application of reference star differential imaging to simultaneously subtract the coronagraphic image of a primary star (\acenA) and the point spread function (PSF) of its binary companion (\acenB) to conduct a deep search for exoplanets and exozodiacal dust emission. We achieve a typical 5$\sigma$ point source contrast sensitivity between $\sim$$10^{-5}$--$10^{-4}$ at separations $\gtrsim$~1\arcsec\ and an exozodiacal disk (coplanar with \acenAB) sensitivity of \added{$\sim$5--8$\times$} the Solar System's zodiacal cloud around \acenA. The latter is an extraordinary limit, representing the deepest sensitivity to exozodiacal disks achieved for any stellar system to date. Additionally, post-processing with the principal component analysis-based Karhunen-Lo\'eve Image-Processing algorithm reveals a point source, called \objS, in August 2024, detected at S/N~$=$~4--6 (3.3--4.3$\sigma$), a projected separation of $\approx$1\farcs5 (2 au), and with a F1550C flux density (contrast) of $\approx$3.5 mJy ($\approx 5.5 \times 10^{-5}$). Various tests conducted with the available data show that \objS is unlikely to be a detector artifact or PSF subtraction artifact and confirm that it is neither a background nor a foreground object. \objS is not re-detected in two follow-up observations (February and April 2025). \added{If \objS is astrophysical in nature, the only explanation is that it} has moved to a region of poor sensitivity due to orbital motion. We perform PSF injection-recovery tests and provide 2D sensitivity maps for each epoch to enable orbital completeness calculations. Additional observations, with JWST or upcoming facilities, are necessary to re-detect candidate \objS and confirm its nature as a planet orbiting our nearest solar-type neighbor, \acenA. More broadly, this program highlights the complexity of analyzing a dynamic binary astrophysical scene and the challenges associated with confirming short-period ($\sim$few years) planet candidates identified without prior orbital constraints in direct imaging searches. This paper is second in a series of two papers: Paper I (Beichman \& Sanghi et al. 2025, in press) discusses the observation strategy and presents the astrophysical case (physical and orbital properties) for \objS as a planet candidate. 
\end{abstract}


% \tableofcontents
\section{Introduction}
At a distance of 1.34 pc from Earth, $\alpha$ Centauri is the Solar System's nearest known neighbor. This stellar system has three components in a hierarchical orbital configuration: $\alpha$~Centauri~A (\acenA, Rigil Kentaurus) and $\alpha$~Centauri~B (\acenB, Toliman) together form an eccentric close binary \citep[semimajor axis $= 23.3$~au, $e=0.52$;][]{akeson_precision_2021} and are orbited by Proxima Centauri ($\alpha$~Centauri~C) at a distance of $\approx$8200 au. The $\alpha$ Centauri System's age is estimated to be $5.3 \pm 0.3$ Gyr \citep{thevenin_asteroseismology_2002, joyce_classically_2018}, similar to our Solar System's age of $\sim$4.6 Gyr. Among the three components, \acenA \citep[G2V, $M_A = 1.0788\;M_\odot$, $R_A = 1.2175\;R_\odot$;][]{akeson_precision_2021} stands out as the closest Sun-like star to Earth and an exceptional target for exoplanet direct imaging searches. It is $\approx$2.7 times closer and $\approx$3.3 times more luminous than the next most favorable G-type star $\tau$ Ceti. This places the habitable zone of \acenA \citep[defined as the separation of Earth-equivalent insolation, $\approx$1.27 au;][]{turnbull_exocat-1_2015} at a modest angular separation of 0\farcs95 (compared to 0\farcs2 for $\tau$ Ceti), resolvable by current ground- and space-based telescopes.

Presently, there are no confirmed planets around \acenA. While \acenB could be responsible for removing planets from the system, dynamical studies suggest a stable zone exists at separations $\lesssim$3 au around \acenA where planets could survive over long timescales \citep{wiegert_stability_1997, quarles_long-term_2016}. The earliest attempts to image companions around \acenA used the Hubble Space Telescope's Planetary Camera in the near-infrared at 1.02 $\mu$m \citep{schroeder_searching_1996, schroeder_search_2000}. These observations were sensitive to 5 Gyr old brown dwarfs with masses $>$~40~\mj\ at separations $>$5\arcsec\ from \acenA. More recently, a 100-hour high-contrast imaging (HCI) campaign with the VISIR mid-infrared camera (10--12.5 $\mu$m) on ESO's Very Large Telescope (VLT) as part of the NEAR (New Earths in Alpha Centauri Region) Breakthrough Watch Project detected a candidate object, called \objC, at a separation of $\approx$1.1 au \citep{wagner_imaging_2021}. The source could be a $\gtrsim$300 K planet with a radius $\gtrsim$3.3 \rearth\ or dust in a $\sim$60 Zodi disk. It has not been possible to re-observe \objC to confirm its nature and exclude alternate possibilities such as an instrumental artifact of unknown origin. Additionally, precision radial velocity measurements limit the minimum mass, $M\mathrm{sin}(i)$, of any planet orbiting \acenA with a period $\lesssim$~1000 days to be $\lesssim$ 100~\mearth\ \citep[``red noise" simulations;][]{zhao_planet_2018}.

\begin{figure*}[tb!]
\centering
\includegraphics[width=\textwidth]{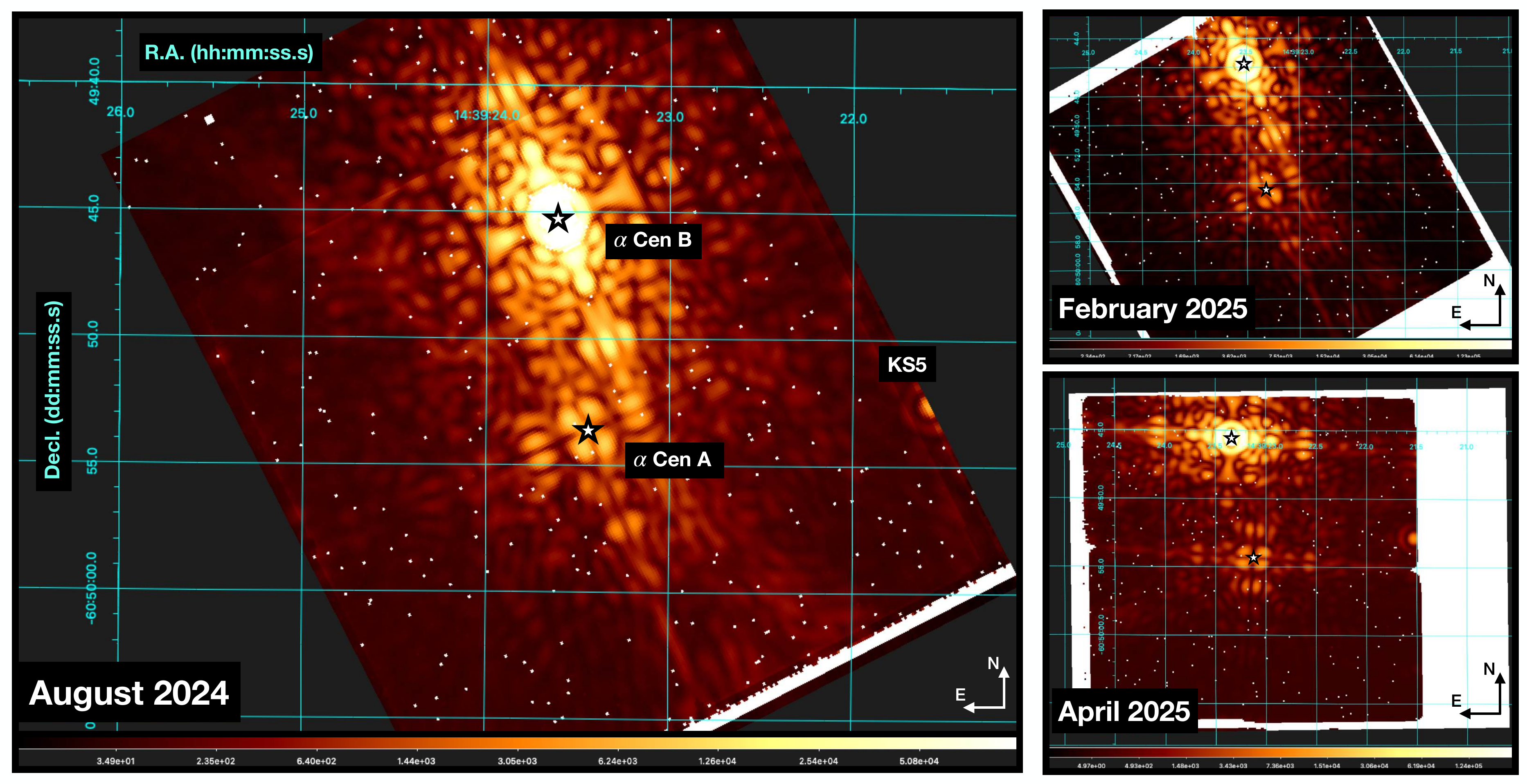}
\caption{\emph{Left:} An ``off-the-shelf" Stage 2b F1550C image of the \acenAB\ system downloaded from MAST for the August 2024 observation (no bad-pixel correction). The image is oriented North up and East left. The white stars denote the approximate positions of \acenB, saturated near the top of the image, and \acenA\ in the lower part of the image placed behind the 4QPM. At the edge of the detector, to the West of \acenA, is a known background source KS5 \citep{kervella_close_2016}. The colorbar is logarithmically scaled in units of MJy/sr. \emph{Right:} Same as the left image for February 2025 (top) and April 2025 (bottom).}
\label{fig:fullfr}
\end{figure*}

The launch of the \emph{James Webb} Space Telescope (JWST) opens up new avenues for direct imaging planet and disk searches around \acenA \citep{beichman_searching_2020, sanghi_preliminary_2025}. Pre-flight simulations by \citet{beichman_searching_2020} showed that the JWST Mid-InfraRed Instrument (MIRI) coronagraph \citep{boccaletti_mid-infrared_2015, wright_mid-infrared_2015} at 15.5 $\mu$m can image planets as small as 5~\rearth\ at separations of 1--3~au including, potentially, the VLT/NEAR \objC candidate. \added{Additionally, with its ability to resolve \acenA's habitable zone, JWST/MIRI could also detect an exozodiacal dust cloud only $\sim$3--5$\times$ the brightness of the Solar System's zodiacal cloud \citep{beichman_searching_2020}.} In this paper, we present results from three epochs of JWST/MIRI coronagraphic imaging observations of \acenA that conducted a deep search for planets and zodiacal dust emission in \acenA's habitable zone. This paper is the second in a series and is preceded by Beichman \& Sanghi et al.~(2025, in press) (also referred to as Paper I). It is organized as follows. Section~\ref{sec:obs} summarizes the observational sequences executed with JWST. Section~\ref{sec:preproc} discusses the initial pre-processing steps for the data. Section~\ref{sec:aug-24-process} describes the stellar point spread function (PSF) modeling and subtraction procedures with the first epoch observations and investigates the nature of a candidate planetary signal. Section~\ref{sec:feb25-proc} presents the results after PSF subtraction for the second and third epoch observations and discusses if the candidate is recovered either as a background object or an orbiting companion. Section~\ref{sec:measure} details the procedures used to estimate the astrometry and photometry of the candidate. Section~\ref{sec:sensitivity} presents the overall sensitivity of our observations to planets around \acenA. Section~\ref{sec:snr_exozodi} discusses our search for extended emission and places new upper limits on its presence around \acenA.  Finally, Section~\ref{sec:conclusion} summarizes our conclusions. \added{Appendix~\ref{sec:app-bin} demonstrates the benefits of using all reference integrations without binning in PSF subtraction} and Appendix~\ref{sec:recover-back} presents the recovery of a known, faint, background source in the individual epoch datasets.

\section{Summary of Observations}
\label{sec:obs}
Here, we provide a summary of JWST observations of the nearest solar-type star, \acenA. For a complete description, we refer the reader to Paper I (Beichman \& Sanghi et al.~2025, in press). JWST successfully observed \acenA through the MIRI coronagraph in the F1550C (15.5~$\mu$m) filter on three separate occasions: August 2024 (Cycle 1 GO, PID \#1618; PI: Beichman, Co-PI: Mawet), February 2025 (Cycle 3 DDT, PID \#6797; PI: Beichman, Co-PI: Sanghi), and April 2025 (Cycle 3 DDT, PID \#9252; PI: Beichman, Co-PI: Sanghi). The observing set-up and strategy was identical for all three observation epochs.

\subsection{Executed Sequences}
The observations were designed to acquire two-roll angle sequences of \acenA to improve sensitivity with reference star differential imaging \citep[RDI:][]{lafreniere_hstnicmos_2009} by (1)~providing a means to de-correlate diffraction features from \acenA and \acenB, and (2)~modulating the transmission of the coronagraph in case a source falls near the coronagraph transition boundary for one of the rolls. It also enables angular differential imaging \citep[ADI:][]{liu_substructure_2004, marois_angular_2006}, however, significant gains compared to RDI are not expected at small angular separations \citep[e.g.,][]{carter_jwst_2023}. The total duration of the exposure per roll angle was 9287.36 seconds comprising of 1250 integrations. However, due to guide star failures, only one roll angle observation was obtained both in August 2024 and February 2025. A full two-roll angle sequence (9$^\circ$ roll) was obtained in April 2025. 

The reference star for these observations, \emus, was observed twice, bracketing the two roll angle observation of \acenA, with the MIRI coronagraph (F1550C) in a 9-point small grid dither (SGD) pattern. This enables us to assemble a coronagraphic image library with a diversity of stellar positions behind the Four Quadrant Phase Mask (4QPM) for RDI and maximize the chance of obtaining a well-matched PSF reference for \acenA. The total duration of the exposure, per dither position, was 2971.792 seconds comprising of 400 integrations. One of the SGD sequences in August 2024 suffered from a guide star failure. Both SGD sequences were successful in February and April 2025.

\begin{figure*}[htb!]
\centering
\includegraphics[width=\textwidth]{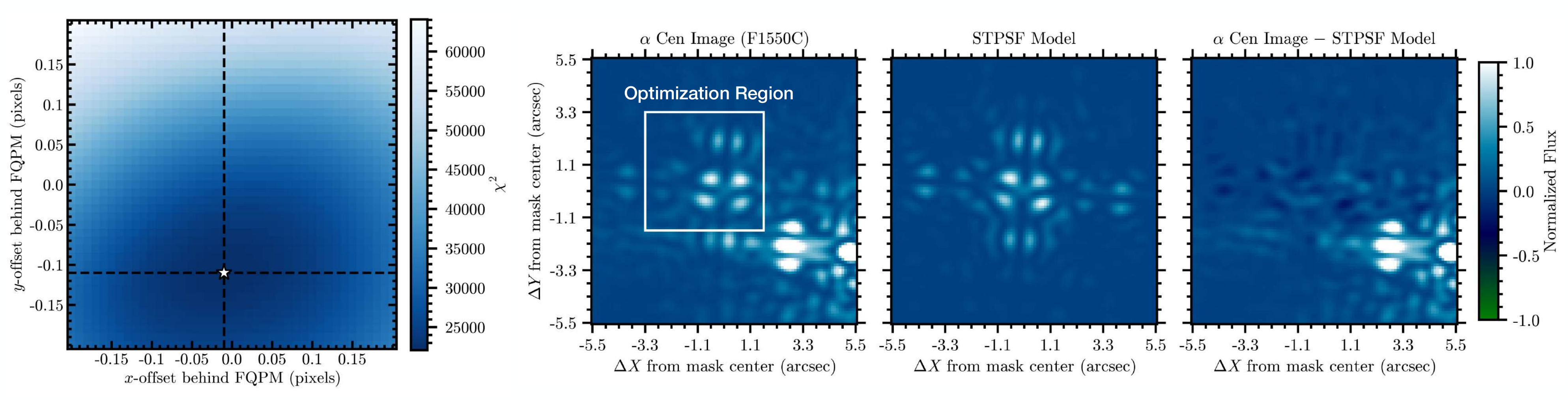}
\caption{Example determination of stellar position behind the MIRI Four Quadrant Phase Mask (FQPM/4QPM) for \acenA in the August 2024 observation. \emph{Left:} $\chi^2$ map obtained after subtracting \webbpsf models generated by varying the stellar position behind the mask (marginalized over the model PSF flux scale factor parameter). The best-fitting PSF (minimum $\chi^2$) is marked with a white star. \emph{Right:} from left to right, the observed \acenA PSF, best-fitting \webbpsf model, and residual image after taking the difference are shown. The white box designates the region over which $\chi^2$ was minimized.}
\label{fig:mask-position}
\end{figure*}

We also placed the reference star, \emus, at the off-axis position of \acenB on the MIRI detector and observed it with the coronagraph in the optical path (F1550C). This enables us to obtain accurate reference images for the removal of \acenB's PSF in post-processing\footnote{\acenA was also observed in July 2024 to test the offset strategy for target acquisition. However, no \emus reference at the detector position of \acenB was acquired. Hence, this dataset was not found to be suitable for a planet or exozodi search.}. Note that the detector position of \acenB changes between observations at the two roll angles. Thus, two distinct off-axis \emus reference observations were required. The total duration of the exposure per observation was 9287.36 seconds comprising of 1250 integrations. All planned off-axis \emus reference observations were successful across the three observation dates.

\added{Finally, to mitigate the effects of the MIRI ``glowstick" \citep{boccaletti_jwstmiri_2022}, we obtained dedicated backgrounds for each science observation (matching detector setups). The position of the background observation was chosen such that it appeared relatively blank in Spitzer or WISE images. Background images were acquired immediately after each \acen roll and \emus on-axis reference observation (the off-axis \emus observations have the same detector setup as the \acen rolls and the corresponding background images are re-used). Each background field was observed twice (for a given observation) with 5\arcsec\ shifts in the center position to help mitigate the effects of sources in the fields.}

\subsection{An Unusual HCI Astrophysical Scene}
The \acenAB system is a complex and dynamic astrophysical scene (Figure~\ref{fig:fullfr}), distinct from those tackled by traditional direct imaging programs, that is both challenging for MIRI observations (as discussed in Paper I) and data analysis. First, the position of \acenA behind the MIRI 4QPM is inconsistent between observation epochs (and even between two rolls acquired in the same epoch) at the $\sim$10~mas level (\S\ref{sec:mask-pos}) because of uncertainties in the blind offset target acquisition procedure (see Paper~I). This changes the coronagraphic image of \acenA for every observation. Second, a deep search for planets and exozodiacal emission not only requires the subtraction of residual starlight from \acenA behind the MIRI coronagraph, but also from its binary companion. \acenB is observed $\sim$7\arcsec--8\arcsec\ away at full brightness on the MIRI detector, unocculted, and with its PSF diffraction features superimposed on those from \acenA's coronagraphic image in the region of interest for planet and disk searches. Third, \acenB's position on the detector and PSF orientation both change for every observation epoch (Figure~\ref{fig:fullfr}) due to orbital motion and the exact V3PA angle of observation, respectively. This, for example, restricts us to only using reference frames acquired in the same epoch for post-processing and affects the achieved sensitivity (as discussed in \S\ref{sec:sensitivity}). Thus, \acenAB is a unique system for high contrast imaging data reduction. \added{In this paper, we develop novel techniques to analyze such datasets.}

\begin{figure*}[htb!]
\centering
\includegraphics[width=\textwidth]{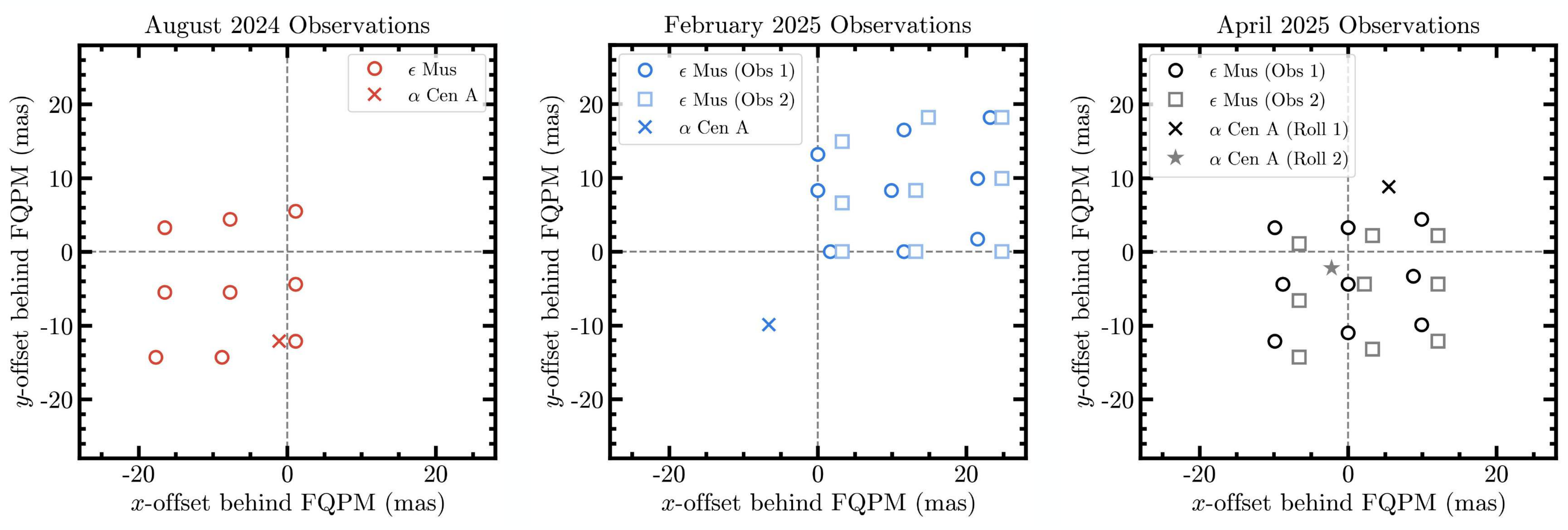}
\caption{Estimated positions of \emus and \acenA behind the MIRI Four Quadrant Phase Mask (FQPM/4QPM) in the 9-point small-grid-dither (SGD) and single/two roll observations, respectively, for each observation epoch (with respect to mask center). The August 2024 and April 2025 (Roll 2 only) \acenA observations have \emus reference images well matched in terms of position behind the 4QPM.}
\label{fig:dither-map}
\end{figure*}

\section{Data Pre-processing}
\label{sec:preproc}

\subsection{JWST Pipeline Processing}
\label{sec:jwst-pipeline}
We obtained the raw (*uncal.fits) images from the Mikulski Archive for Space Telescopes (MAST). The images were pre-processed using \texttt{spaceKLIP} \citep{kammerer_performance_2022, carter_jwst_2023, carter_spaceklip_2025}, a community-developed repository that wraps key \texttt{jwst} \citep{bushouse_jwst_2025} pipeline steps with modifications and additions appropriate for high contrast imaging. \added{The reduction used the \texttt{jwst} pipeline version 1.18.0, the CAlibration REference Data System (CRDS) version 12.1.7, and the CRDS context file \texttt{jwst\_1364.pmap}.} We skipped dark current subtraction, to avoid introducing additional noise due to the low signal-to-noise ratio of the available darks. The data were read ``up the ramp" using the \texttt{LIKELY} algorithm described in \citet{brandt_optimal_2024} and corrected for jumps using a threshold of 8 \citep[following previous work with MIRI coronagraphy;][]{boccaletti_jwstmiri_2022, carter_jwst_2023, malin_unveiling_2024}. We skipped the flat-fielding ``stage 2" calibration step, to avoid biasing photometry of sources near the transitions of the 4QPM following \citet{malin_unveiling_2024, malin_first_2024}.

The Stage 2 (*calints.fits) images were iteratively corrected for bad pixels (cosmic rays, hot pixels, dead pixels, etc). First, pixels flagged during the ramp fitting stage, and by the \texttt{jwst} pipeline during Stage 2, were filled in with an interpolating spline with a kernel size of 3 pixels. Then, temporal outliers (between integrations in a given pointing) were flagged using sigma-clipping with a threshold of 6, and replaced with their temporal median. Finally, additional spatial outliers were flagged using sigma-clipping with a threshold of 3 and a search box of 5 pixels on a side, and replaced with an interpolating spline with a kernel size of 3 pixels, as in the first step. Once corrected for bad pixels, the associated MIRI 4QPM background exposures were subtracted from the \acenAB science, \emus on-axis reference, and \emus off-axis reference exposures that matched their readout pattern.

\begin{figure*}[tb!]
\centering
\includegraphics[width=\textwidth]{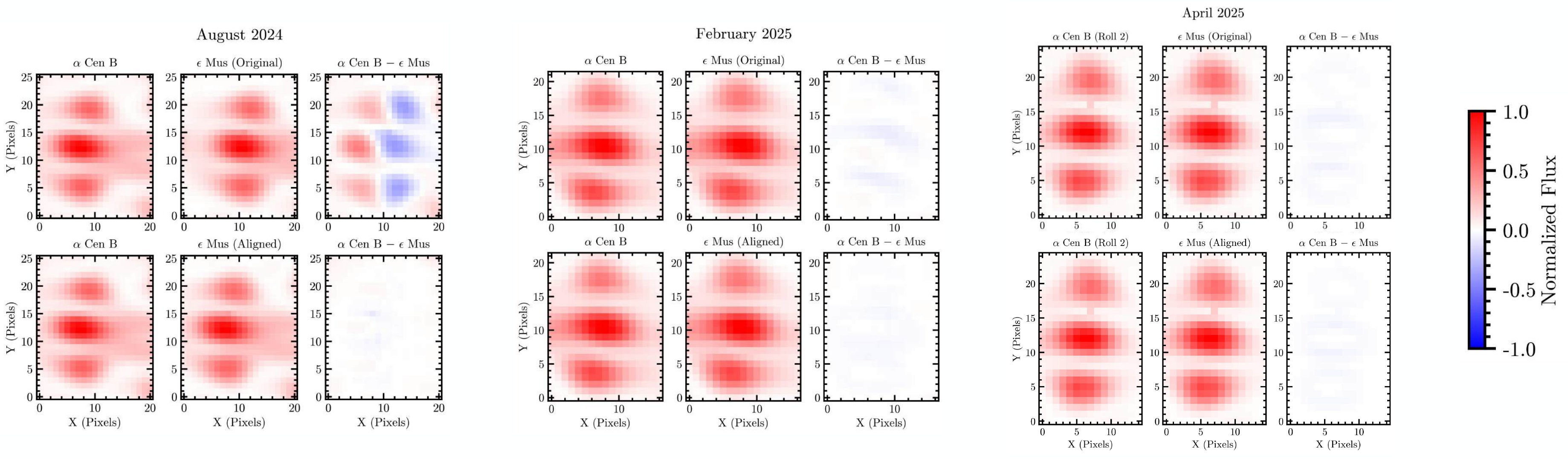}
\caption{Alignment of the off-axis \emus reference frames to the corresponding \acenB observations using a bright diffraction feature in each of the epochs (for April 2025, only the roll 2 observation is shown). \emph{Top panel:} for a given epoch, from left to right, the \acenB observation, \emus reference frame, and difference image. \emph{Bottom panel:} similar to the top panel but with the \emus reference frame aligned to the \acenB observation.}
\label{fig:align-emus}
\end{figure*}

\subsection{Stellar Positions Behind the MIRI/4QPM}
\label{sec:mask-pos}
Estimating the position of \emus and \acenA behind the MIRI/4QPM enables (1)~verification of the accuracy of target acquisition with blind offsets from background stars; (2)~determination of the pointing stability between observations in the 9-point small-grid-dither (SGD) observation of \emus; and (3)~identification of the \emus reference observation that best matches the \acenA PSF. Broadly, we find the stellar position behind the MIRI/4QPM by minimizing the $\chi^2$ on comparison with a grid of \webbpsf models \citep{perrin_simulating_2012, perrin_updated_2014} generated for varying offsets. The detailed steps are described below.

We begin with the background-subtracted Stage 2b data products obtained following the processing steps described in Section~\ref{sec:jwst-pipeline}. For \emus, we average the 400 integrations at each dither position and for \acenA, the 1250 integrations at the single roll to obtain mean frames for comparison with \webbpsf models. The uncertainty estimate for each pixel in a given integration is retrieved from the \texttt{ERR} extension in the \texttt{calints} data product and appropriately combined to obtain the per pixel uncertainty for the mean frames. Next, we define the region over which $\chi^2$ is optimized. For \emus, we select a $90 \times 90$ pixel subarray centered on the coronagraph center as retrieved from \texttt{CRPIX1} and \texttt{CRPIX2} header values. For \acenA, we select a $45 \times 45$ pixel subarray encompassing the central PSF but offset to exclude the majority of the two quadrants where there is significant contamination from the unocculted off-axis PSF of \acenB (example in Figure~\ref{fig:mask-position}). \webbpsf models are computed for a grid of pointing offsets between $-$0.2 and 0.2 pixels ($-$22 mas to 22 mas) in steps of 0.01 pixels (1.1 mas) along each axis. The models include detector effects \added{(interpixel capacitance and charge diffusion, as discussed in \url{https://stpsf.readthedocs.io/en/latest/jwst_detector_effects.html})}, are generated using the closest-in-time on-sky optical path difference (OPD) map to the observations, and oversampled by a factor of 4. 

The fitting procedure registers the \webbpsf model to the mean comparison frame using a Fourier image shift \citep[][]{greenbaum_first_2023} implemented with the \texttt{webbpsf\_ext} package \citep{leisenring_webbpsf_2025}. The required shift is determined with the \texttt{image\_registration} package’s \texttt{chi2\_shift} function and is applied to the oversampled model PSF before binning down to the native detector sampling. We compute $\chi^2$ over all pixels in the optimization region for the grid of stellar positions behind the mask as well as a varying flux scale factor for the model PSFs. A noise floor was applied to the pixels in the optimization regions to ensure that the pixels belonging to the image background do not artificially inflate the $\chi^2$ value. We selected the smallest uncertainty value among pixels that were a part of prominent features of the stellar diffraction pattern in the optimization region for the same. A clear $\chi^2$ minimum is identified for all the \acenA observations and all the \emus observations (one example shown in Figure~\ref{fig:mask-position}). The typical uncertainty in our estimate of the stellar position behind the mask is 1.1 mas.

Our estimates of the stellar position behind the mask for the \emus and \acenA observations indicate that the initial coronagraph pointing after blind offsets is accurate to $\sim$10 mas along each axis (Figure~\ref{fig:dither-map}), consistent with the typical pointing accuracy of MIRI coronagraphic observations \citep{boccaletti_jwstmiri_2022, rigby_science_2023}. The 9-point SGD pattern was effective at providing a reference PSF that closely matches the August 2024 science observation (Figure~\ref{fig:dither-map}). However, due to a significant pointing mismatch ($\approx$25 mas between \acenA and the SGD center), none of the dithered reference positions closely match that of the February 2025 science observation. For the April 2025 science observations, the SGD was effective at providing reference PSFs bracketing the position of \acenA behind the 4QPM for roll 2, but not for roll 1.

\subsection{Off-axis \emus Reference Image Alignment}
\label{sec:align}
The off-axis \emus observations serve as a reference for the subtraction of \acenB's PSF. First, we need to align the off-axis \emus reference frames to their corresponding science observation for each observation epoch. The images are registered over a small subarray centered on a bright \acenB diffraction feature using an interpolation-based image shift implemented with the \texttt{pyklip.klip.align\_and\_scale} function \citep{wang_pyklip_2015}. We did not use Fourier image shifts as they introduced Gibbs artifacts in the shifted images from the saturated regions of \acenB's PSF. The diffraction feature was chosen such that it avoids saturated regions of the \acenB PSF and shows a structure distinct from \acenA's features. The required shift is determined with the \texttt{image\_registration} package’s \texttt{chi2\_shift} function (Figure~\ref{fig:align-emus}). We find that an ($x$, $y$) shift of ($-$3.00, $-$0.10) pixels is needed to align the \emus reference PSF to \acenB's PSF in the August 2024 observations. Similarly, for the February 2025 observation we require an ($x$, $y$) shift of ($-$0.10, $-$0.10) pixels. For roll 1 and 2 of the April 2025 observation, the shift required to align the associated off-axis \emus reference PSF is (0.03, 0.06) pixels and (0.03, $-$0.04) pixels, respectively.

\begin{figure*}[htb!]
\centering
\includegraphics[width=\textwidth]{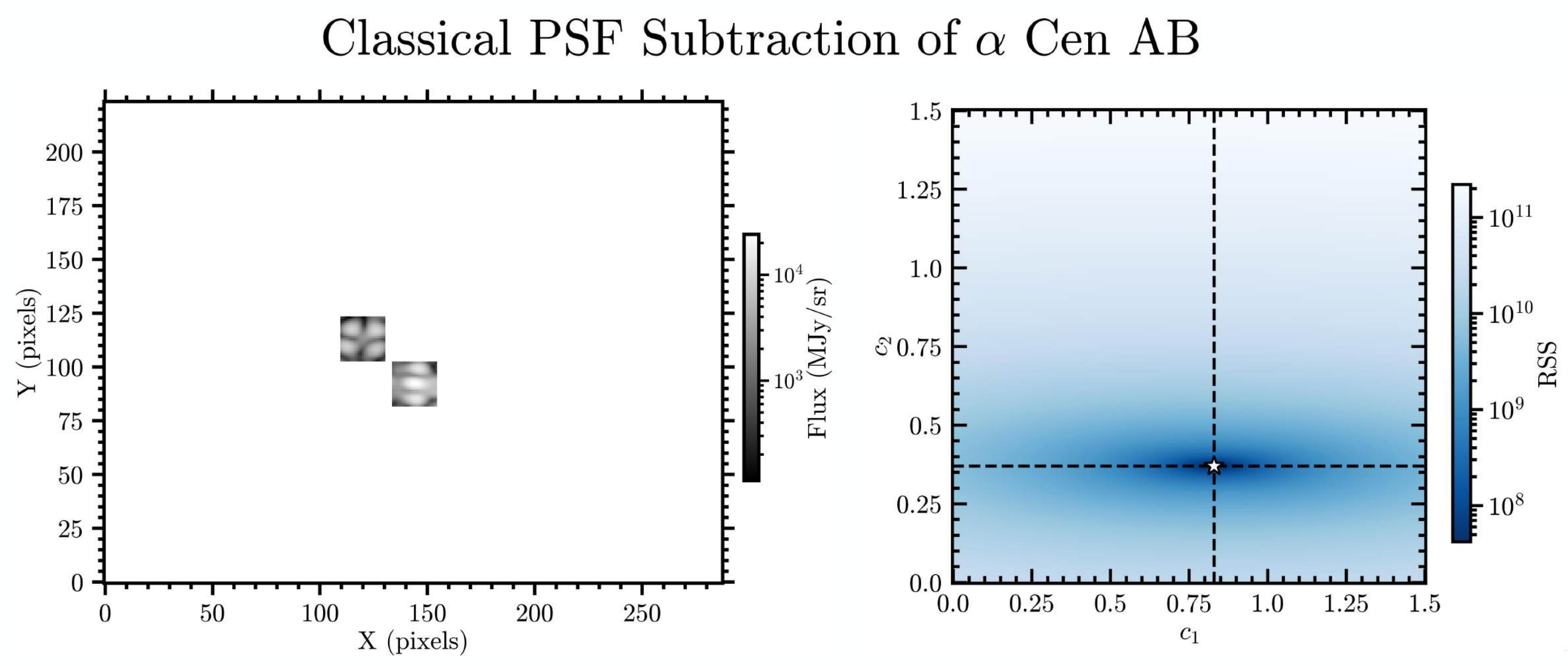}
\caption{Subtraction of \acenAB using a linear combination of reference images. \emph{Left:} Full \acenAB detector image showing the regions over which the residual sum of squares (RSS) was minimized to obtain coefficients $c_1$ (for \acenA's PSF model) and $c_2$ (for \acenB's PSF model). \emph{Right:} RSS map obtained after subtracting the model PSFs. The best-fitting (minimum RSS) coefficient pair ($c_1=0.830$, $c_2=0.370$) is marked with a white star.}
\label{fig:linear-comb}
\end{figure*}

\begin{figure*}[htb!]
\centering
\includegraphics[width=\textwidth]{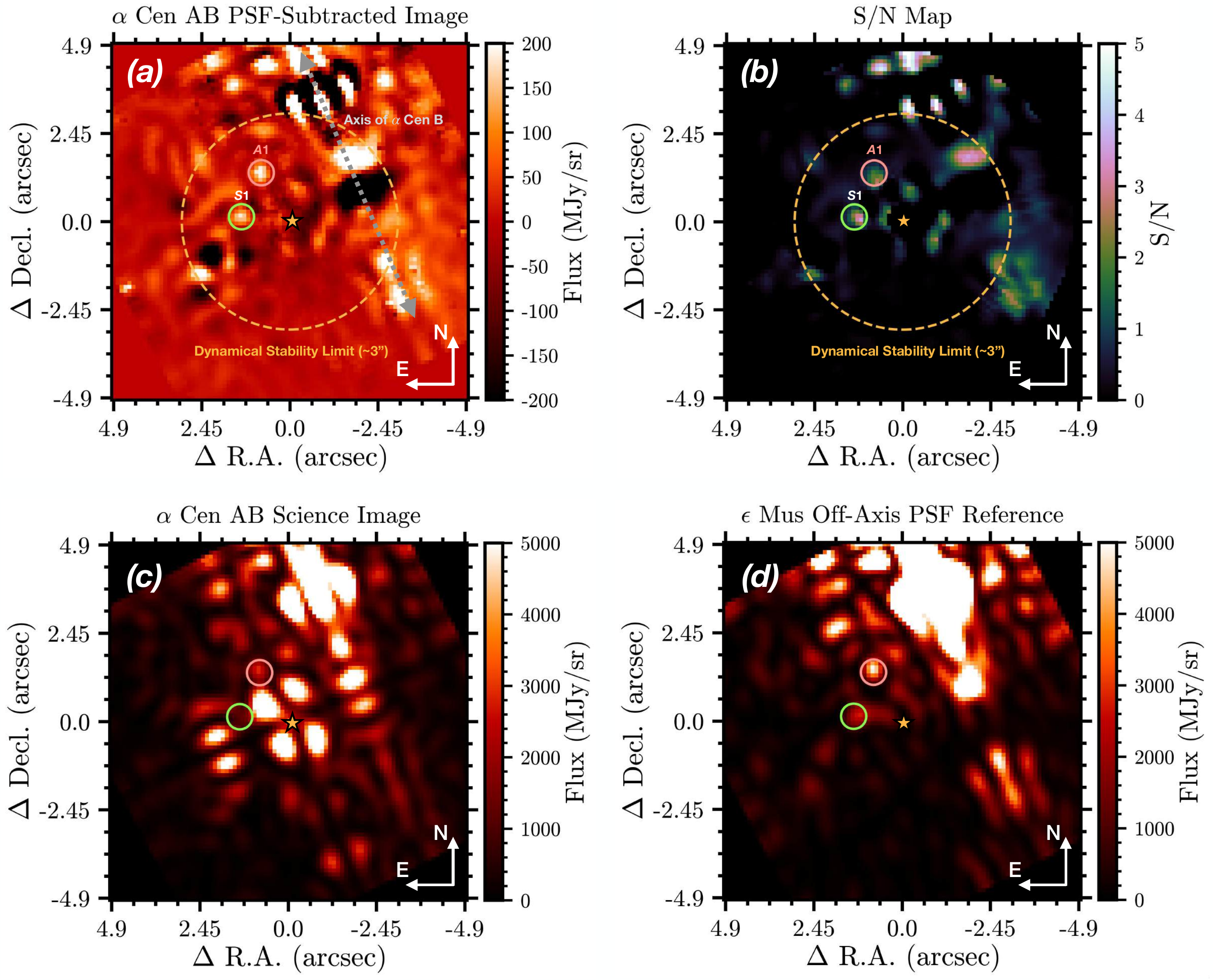}
\caption{Results from classical PSF subtraction of \acenAB \added{for the August 2024 observations}. \emph{Top panel:} The de-rotated residual image (a; North up, East left) and associated S/N map (b) after subtraction of \acenAB with a linear combination of reference images, using the best-fit $c_1$ and $c_2$ coefficient values. Visually identified point sources, \objS and \emph{A}1, are circled. The star symbol marks the center of \acenA. \emph{Bottom panel:} The left image (c) is the de-rotated \acenAB science integration before PSF subtraction and the right image (d) is the de-rotated off-axis \emus reference for \acenB. The latter acts as a proxy for and shows the diffraction features expected from \acenB in vicinity of \acenA. A red circle marks the position of \emph{A}1, which is co-located with a diffraction feature of \acenB's PSF and likely a residual speckle artifact. A green circle marks the position of \objS, which may be associated with an \acenB diffraction feature but is the only source we confidently re-detect in the more robust PCA-KLIP reductions.}
\label{fig:linear-comb-results}
\end{figure*}

\begin{figure*}[htb!]
\centering
\includegraphics[width=\textwidth]{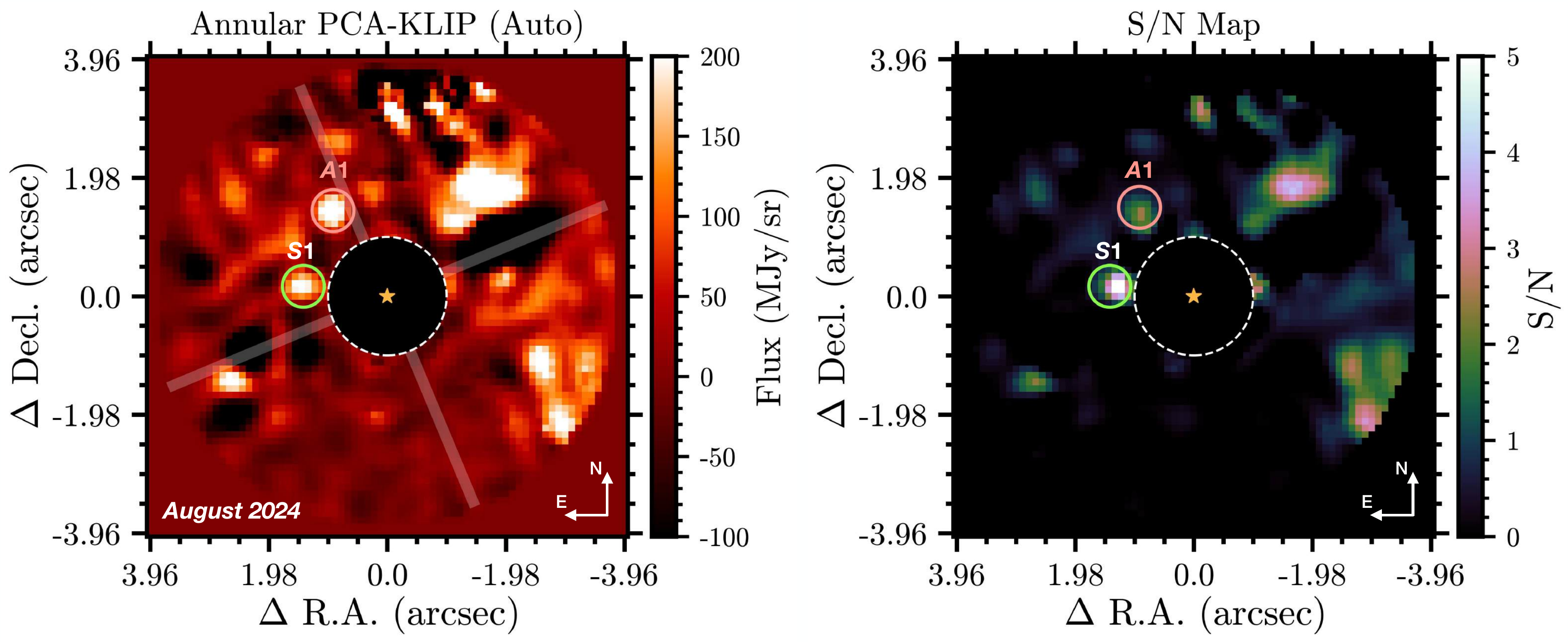}
\caption{Joint PCA-KLIP subtraction of \acenAB for the August 2024 observations with automatic principal component selection. \emph{Left:} The de-rotated residual image (North up, East left) after PSF subtraction in three concentric annuli. The MIRI/4QPM transition boundaries are shown as gray shaded regions. Point source detection \objS and artifact \emph{A}1 (discussed in \S\ref{sec:pcs-vary}) are circled. \emph{Right:} S/N map for the reduction shown in the left panel. \objS is detected with a peak S/N of 5.4 (4$\sigma$). \added{\emph{A}1 is brighter than \objS but detected a lower peak S/N of 2.5 (the residual noise estimated using non-overlapping apertures at the same separation but different position angles, following \citet{mawet_fundamental_2014}, is higher at the separation of \emph{A}1.)}}
\label{fig:joint-pca}
\end{figure*}

\section{August 2024 Data Post-processing}
\label{sec:aug-24-process}
In this section, our goal is to conduct a comprehensive search for point sources representing planetary companions around \acenA, within a radial field-of-view $<$5\arcsec, in the August 2024 observations. According to dynamical stability arguments \citep[e.g.,][]{wiegert_stability_1997}, planets could exist within approximately 4~au ($\approx$3\arcsec) of \acenA . We employ RDI to subtract both \acenA and \acenB and explore two distinct PSF subtraction strategies to ensure robustness of any detections, as detailed below. The failure of the second roll observation of \acenA, in August 2024, precludes the use of ADI. The PSF-subtracted images are visually-inspected for point sources. A corresponding signal-to-noise ratio (S/N or SNR) map is generated for each reduction following the definition in \citet{mawet_fundamental_2014} using the \texttt{snrmap} function in \texttt{vip\_hci} \citep{gomez_gonzalez_vip_2017, christiaens_vip_2023} and is used to assess the strength of any candidates identified.

\subsection{Classical PSF Subtraction: Median-RDI}
\label{sec:linear-comb}
The 1250 \acenAB images are median-combined to produce a single science image for the PSF subtraction procedure. We use the median-combined $7^{\rm th}$ dither observation of \emus ($\rm PSF_{\epsilon,\;\rm on}$) as the reference model for \acenA's PSF due to its similarity in position behind the MIRI/4QPM (see \S\ref{sec:mask-pos}) and the median-combined aligned off-axis observation of \emus ($\rm PSF_{\epsilon,\;\rm off}$) as the reference model for \acenB's PSF. The combined PSF model for \acenAB is then constructed as the linear combination $c_1 \cdot {\rm PSF_{\epsilon,\;{\rm on}}} + c_2 \cdot \rm PSF_{\epsilon,\;\rm off}$. The optimization region for PSF subtraction is chosen to include bright diffraction features from both \acenA and \acenB and is visualized in Figure \ref{fig:linear-comb}. This was inspired by the zone selection methodology of the Locally Optimized Combination of Images \citep[LOCI:][]{lafreniere_new_2007} algorithm. We find that the coefficient pair ($c_1 = 0.830$, $c_2 = 0.370$) minimizes the residual sum of squares (RSS) in the optimization region (Figure \ref{fig:linear-comb}). We experimented with different sets of optimization regions and found similar values for $c_1$ and $c_2$. The above values are adopted for PSF subtraction below.

PSF subtraction is performed over the full detector image using the reference PSFs with the coefficients determined above and the residual image is de-rotated to align North up and East left. Two point-like sources labeled \emph{S}1 and \emph{A}1 are identified around \acenA within the approximate region of dynamical stability by visual inspection (Figure~\ref{fig:linear-comb-results}a) and for S/N~$>$~2.5. Sources \emph{S}1 and \emph{A}1 are detected at a peak S/N of 3.6 and 2.8 (Figure~\ref{fig:linear-comb-results}b), respectively. \emph{A}1 is directly associated with a bright diffraction feature of the \acenB PSF (Figure~\ref{fig:linear-comb-results}d). \emph{S}1 is next to a diffraction feature of the \acenB PSF and may be associated with it (Figure~\ref{fig:linear-comb-results}d). The following, more robust and sensitive, PCA-KLIP analysis will help distinguish diffraction and/or speckle artifacts from true astrophysical sources. The S/N map shows additional structure in the North-West quadrant, however these lie along the axis of \acenB's PSF.

\subsection{Joint PCA-KLIP Modeling of \acenAB}
\label{sec:joint-pca}
In this approach, we model and subtract the stellar diffraction pattern from \acenA and \acenB using the principal component analysis-based Kahrunen Loeve Image Processing algorithm \citep[PCA-KLIP:][]{soummer_detection_2012}. We use a single reference library consisting of both the on- and off-axis \emus reference frames to construct the PSF model for a given subtraction region\footnote{We also explored sequentially subtracting \acenB, first, followed by \acenA, second, with PCA-KLIP. However, this proved challenging due to over-subtraction in the vicinity of \acenA after performing the first step.}. 

\begin{figure*}[htb!]
\centering
\includegraphics[width=\textwidth]{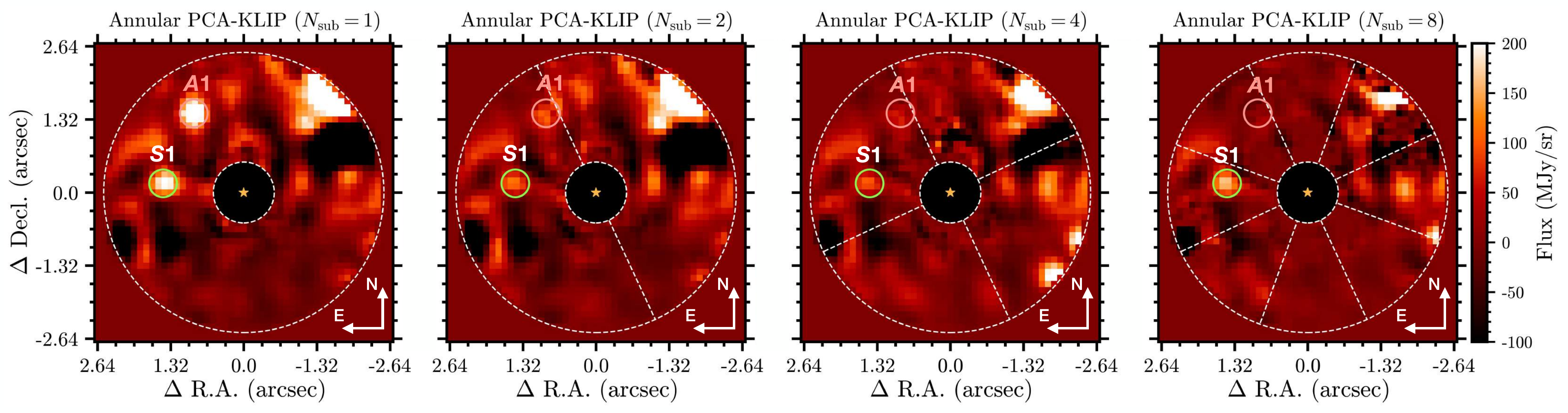}
\caption{Joint PCA-KLIP subtraction of \acenAB \added{for the August 2024 observations} for different number of azimuthal subsections ($N_{\rm sub}$). \objS is recovered in all reductions, whereas \emph{A}1's signal disappears as $N_{\rm sub}$ increases.}
\label{fig:annulus}
\end{figure*}

\subsubsection{PSF Subtraction with Auto-PC Selection}
\label{sec:auto-pc}
We begin by conducting a blind search for astrophysical sources in the vicinity of \acenA. We perform reductions where joint PCA-KLIP is applied to a single annulus with a radial size of 2 FWHM ($\approx1\arcsec$) and centered at varying radial separations between 0\farcs75--3\farcs5 in steps of 0\farcs25 (twelve different reductions). We allow the \texttt{pca\_annulus} function in \texttt{vip\_hci} to auto-select the optimal number of principal components (PCs) for each annulus as the value at which the standard deviation of the residuals (after the subtraction of the PCA approximation) drops below a tolerance value = 0.1. Smaller tolerance values lead to the use of a larger number of PCs. The motivation behind this method of selecting the number of PCs is to optimize the subtraction with respect to the noise while remaining agnostic to possible (a priori, unknown) signals. This strategy works well for RDI reductions, where the PCs do not depend on the astrophysical signal (over-subtraction). However, we note that this is not applicable for ADI reductions, where the PCs do depend on the astrophysical signal (self-subtraction) and optimizing on the noise rather than the signal can significantly impact detectability \citep{pueyo_detection_2016}.

After an extensive search across all our reductions, only one point source was detected at S/N $>$ 5. As a summary, we show a representative post-processed image (and the corresponding S/N map) obtained by combining the reductions conducted in non-overlapping annuli centered at $\left\{1.5\arcsec, 2.5\arcsec, 3.5\arcsec\right\}$ (Figure \ref{fig:joint-pca}). \objS is re-detected at a S/N of 5.4 (corresponding to a 4$\sigma$ detection). The S/N ratio is converted to a Gaussian detection significance level for the equivalent false positive probability using the \texttt{significance} function in \texttt{vip\_hci}. \emph{A}1 is also seen in the reduction but is not detected at significant S/N. It is discussed in more detail in the following section.

\begin{figure*}[htb!]
\centering
\includegraphics[width=\textwidth]{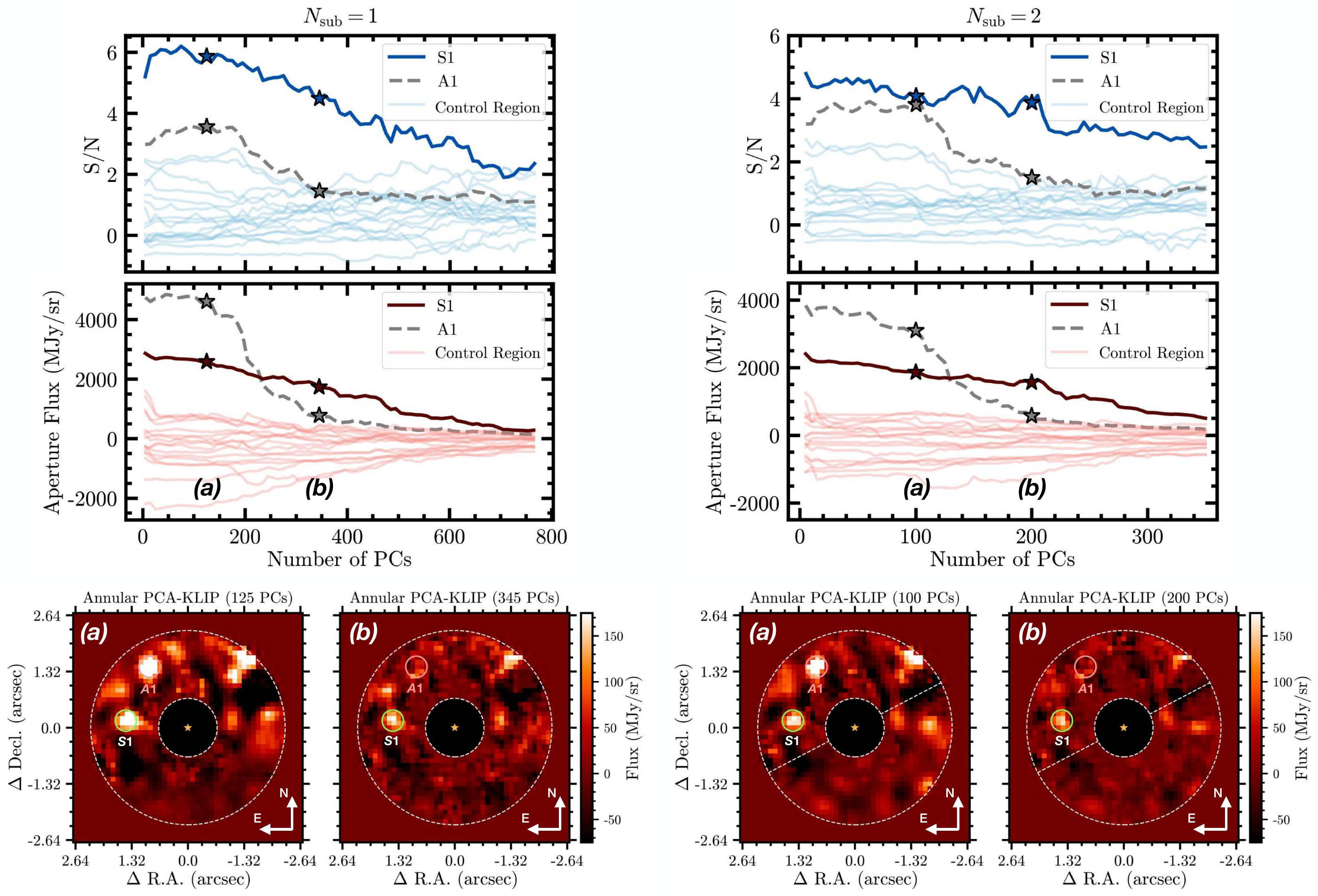}
\caption{Evolution of \objS and \emph{A}1 with the number of principal components (PCs) used in the joint PCA-KLIP subtraction procedure \added{for the August 2024 observations}. The left panel corresponds to reductions where joint PCA-KLIP is applied to a single annulus. The right panel corresponds to reductions where joint PCA-KLIP is applied to a single annulus divided into two subsections. \emph{Top row:} S/N as a function of principal components for \objS and \emph{A}1. They are compared to the S/N of the residual speckle noise estimated in 1 FWHM diameter circular apertures at the same separation as \objS but at different position angles (control region; each semi-transparent line corresponds to an aperture at a different position angle). \emph{Center row:} same as the top row but for the aperture-integrated flux as a function of number of PCs. \emph{Bottom row:} Two representative post-processed images at PCs marked by stars in the above figures. \objS's signal remains stable across PCs, whereas \emph{A}1's signal disappears after a certain number of PCs.}
\label{fig:s1-a1}
\end{figure*}

\begin{figure}[tb!]
\centering
\includegraphics[width=0.45\textwidth]{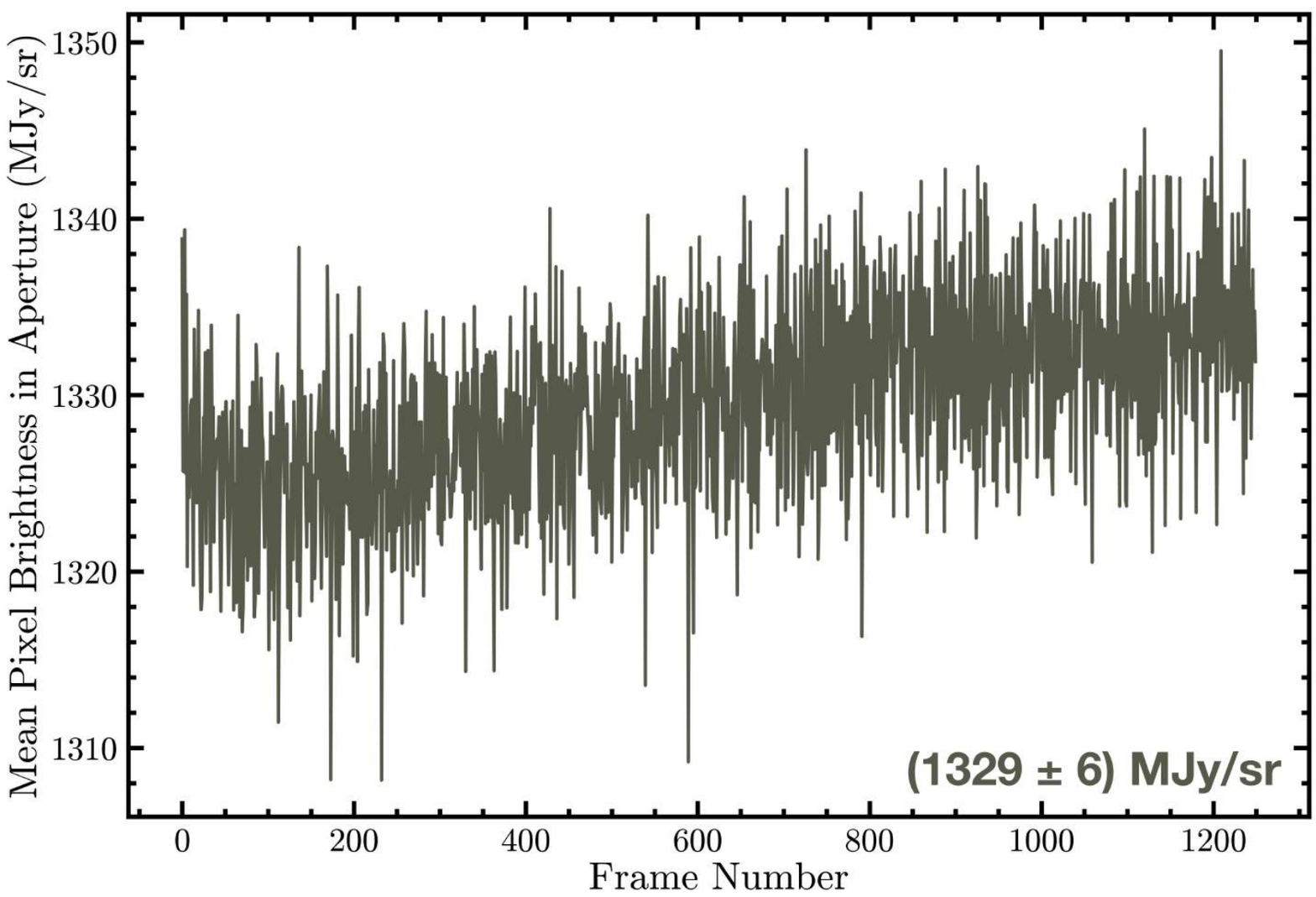}
\caption{The average pixel brightness in a 3 pixel radius aperture centered on \objS in the background-subtracted Stage~2b \acenAB integrations \added{from August 2024}. The time-series does not show any sharp transitions (increases) that would appear as an artifact at the location of \objS after PSF subtraction.}
\label{fig:detector}
\end{figure}

\begin{figure}[tb!]
\centering
\includegraphics[width=0.5\textwidth]{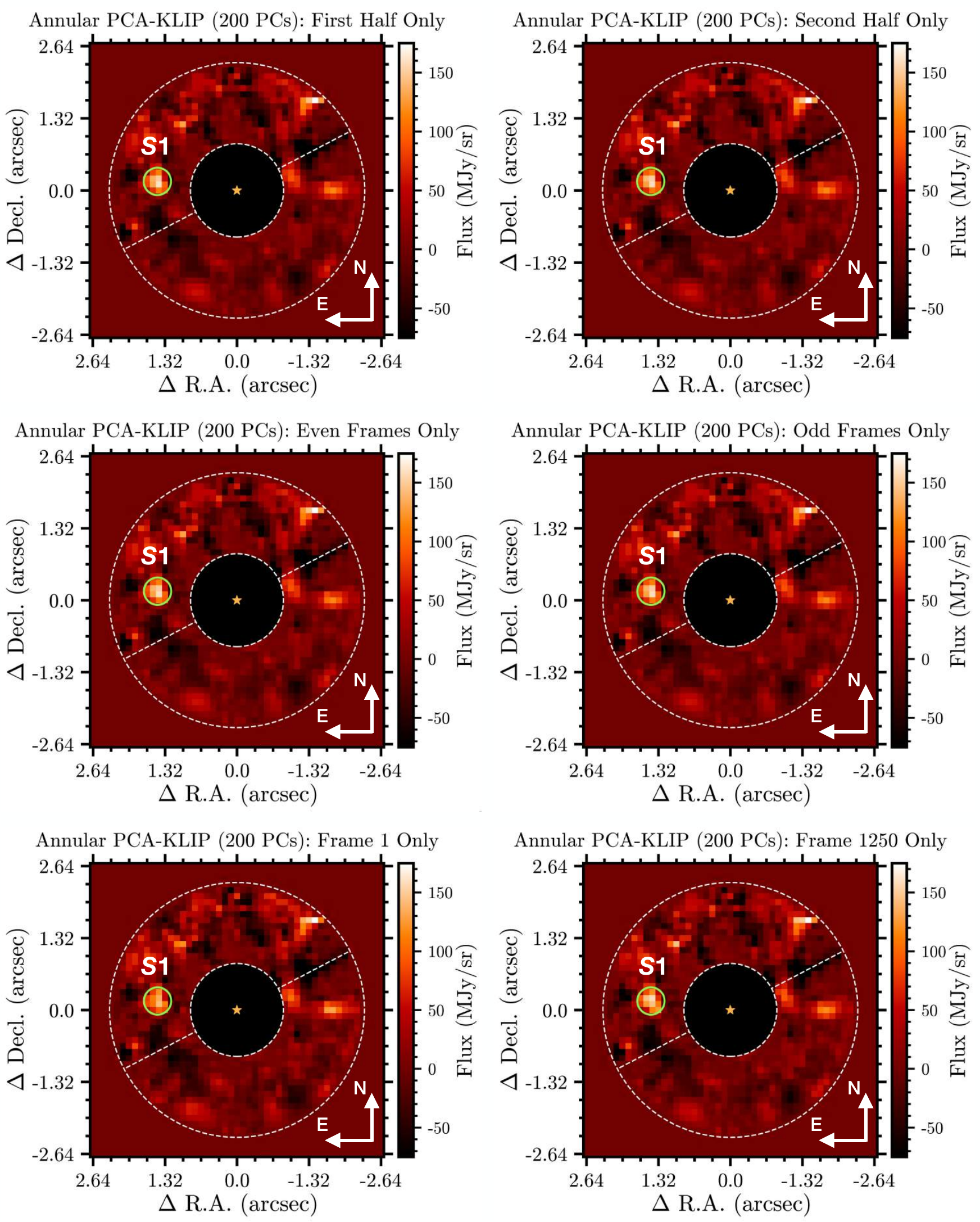}
\caption{Joint PCA-KLIP subtraction of distinct subsets of the \acenAB science integrations \added{from the August 2024 observations}. Reductions performed on the first half, second half, even set, odd set, first frame, and last frame of the integration sequence are shown from top left to bottom right. \objS is consistently detected in all data subsets.}
\label{fig:subset-detections}
\end{figure}

\subsubsection{Robustness of S1 to Varying PSF Subtraction Geometries and Principal Components}
\label{sec:pcs-vary}
PSF subtraction is performed in the annular mode for varying number of subsections ($N_{\rm sub}$), annulus widths, and principal components (PCs) to test the robustness of the \objS candidate. First, we apply joint PCA-KLIP to a 4~FWHM width annulus centered at a radial separation of $\approx$1\farcs5 with no subsections and then divided into two, four, and eight subsections. The number of PCs is automatically chosen by the \texttt{pca\_annular} function as discussed in \S\ref{sec:auto-pc} for each reduction. Second, we apply joint PCA-KLIP to a 3~FWHM width annulus centered at a radial separation of $\approx$1\farcs5 with no subsections and then divided into two subsections along one of the 4QPM transition boundaries. In each case, we perform multiple reductions by varying the number of PCs used incrementally. Across these reductions, we track the S/N and flux (in a 1 FWHM diameter aperture) of \objS and \emph{A}1. We also examine the S/N and aperture flux of the residual speckle noise at the separation of \objS, estimated using non-overlapping 1 FWHM diameter circular apertures at different position angles. 

\begin{figure*}[tb!]
\centering
\includegraphics[width=\textwidth]{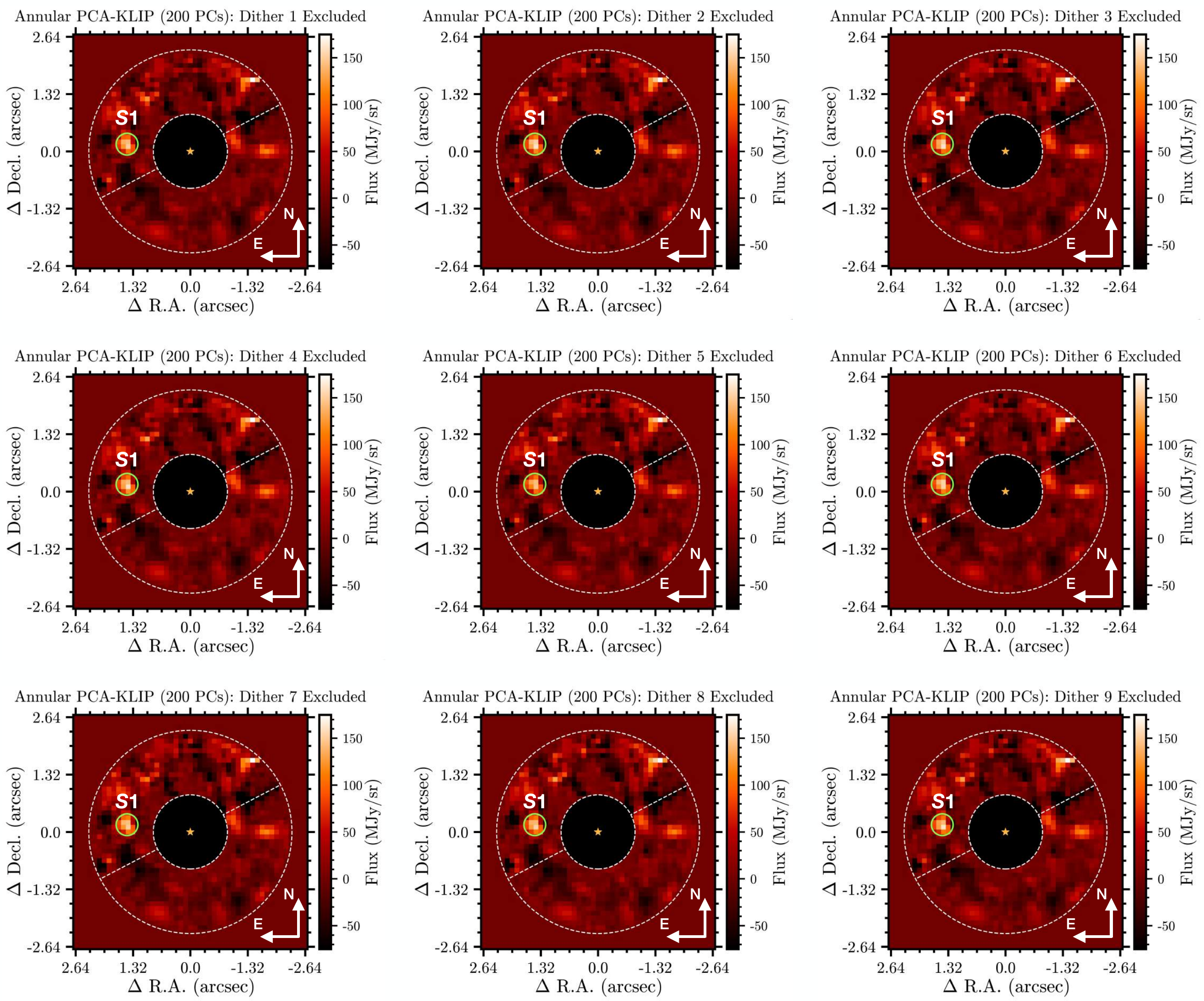}
\caption{Joint PCA-KLIP reductions \added{for the August 2024 observations} where frames at each \emus dither position are iteratively excluded (one at a time) from the reference library. \objS is detected in all reductions indicating that it is not likely an artifact introduced by the \emus coronagraphic images.}
\label{fig:jackknife}
\end{figure*}

Performing the first test, we find that \objS is recovered irrespective of the number of subsections used (Figure~\ref{fig:annulus}). However, \emph{A}1's signal becomes weaker as the number of subsections increases and disappears for the reduction with eight subsections (Figure~\ref{fig:annulus}). Performing the second test, for the two $N_{\rm sub}$ cases, we find that \objS behaves distinctly from both \emph{A}1 and the residual speckle noise. \objS's S/N remains relatively flat at values $>$4 at low PCs and then smoothly decreases as the number of PCs used increases (Figure~\ref{fig:s1-a1}). The aperture flux of \objS exhibits a similar behavior. \emph{A}1's flux, on the other hand, starts out higher than \objS at low PCs but experiences a drastic drop as the number of PCs is increased ($\gtrsim$150 for $N_{\rm sub} = 1$ and $\gtrsim$100 for $N_{\rm sub} = 2$; Figure~\ref{fig:s1-a1}). The S/N of \emph{A}1 shows a similar behavior as its aperture flux. The S/N of the residual speckle noise at the same separation is $<$3 and remains roughly flat across PCs. 

Together, both tests confirm that \emph{A}1 is a residual artifact from \acenB's PSF (as first suspected based on the comparison in Figure~\ref{fig:linear-comb-results}). \emph{A}1 is not robust to changing PSF subtraction geometries and the rapid flux drop above a critical number of PCs is a telltale sign that it corresponds to a PSF feature that was modeled and subtracted out by PCA-KLIP. For \objS, the two tests strengthen the case against it being a speckle or \acenB diffraction artifact. \objS is robust to changing PSF subtraction geometries and the images in Figure~\ref{fig:s1-a1} show that (1)~its flux and S/N evolve smoothly with increasing PCs (the smooth decrease is a feature of over-subtraction with RDI); and (2) it persists as a point source at higher number of PCs after the subtraction of \emph{A}1, which has been directly associated with an \acenB PSF artifact.

\begin{figure*}[htb!]
\centering
\includegraphics[width=\textwidth]{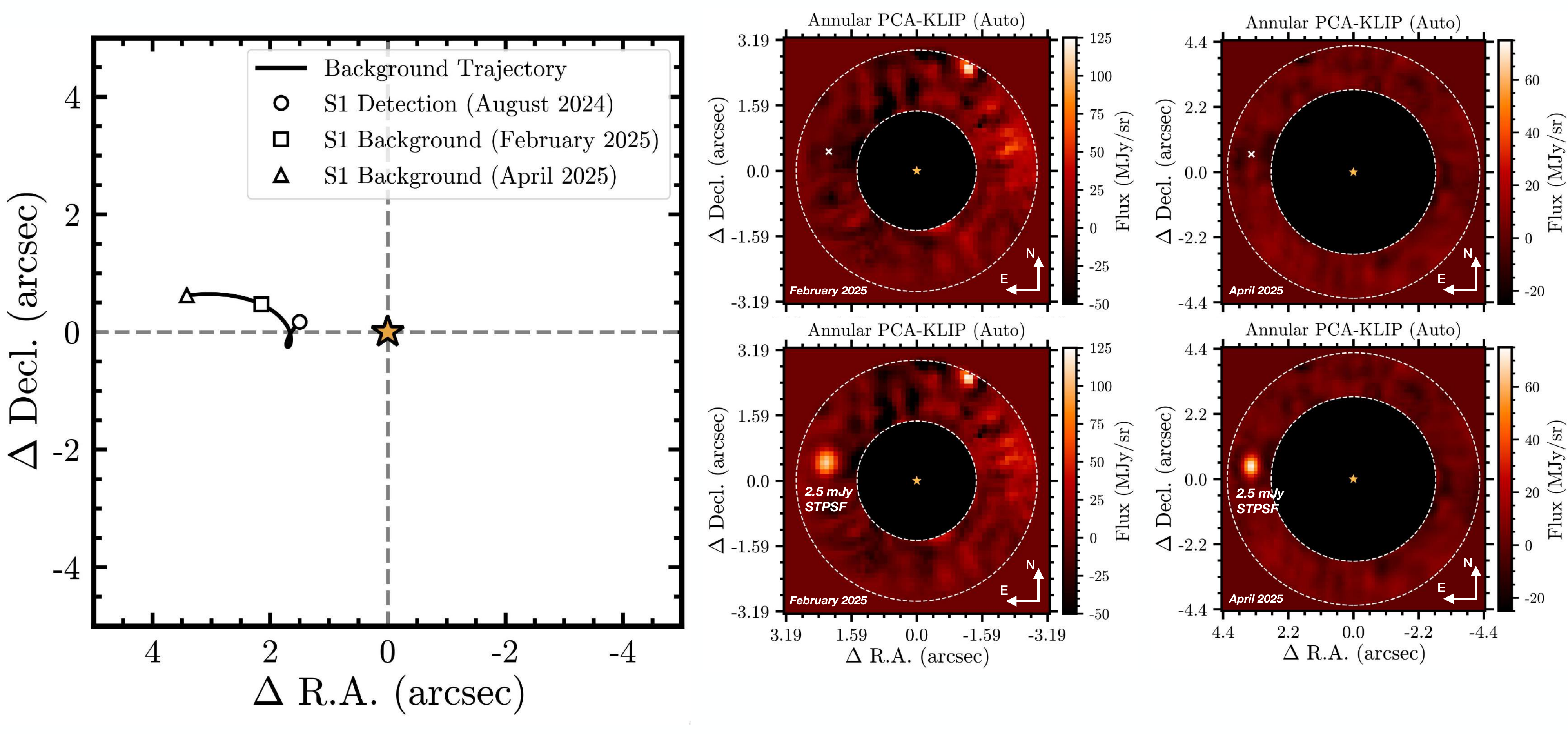}
\caption{\emph{Left:} sky motion of \objS (with respect to \acenA marked with a star symbol), as seen by JWST, assuming it is a stationary background source. The square and triangle symbols mark the expected location of \objS for the February and April 2025 observation dates. \emph{Right:} joint PCA-KLIP reductions for the two epochs showing non-detection of a point source at the expected location for a background object (marked with a cross, top panel). Injection and recovery of a 2.5 mJy \webbpsf model (bottom panel) confirms the sensitivity of the observations to \objS. \objS is not a stationary background object.}
\label{fig:background}
\end{figure*}

\begin{figure*}[htb!]
\centering
\includegraphics[width=\textwidth]{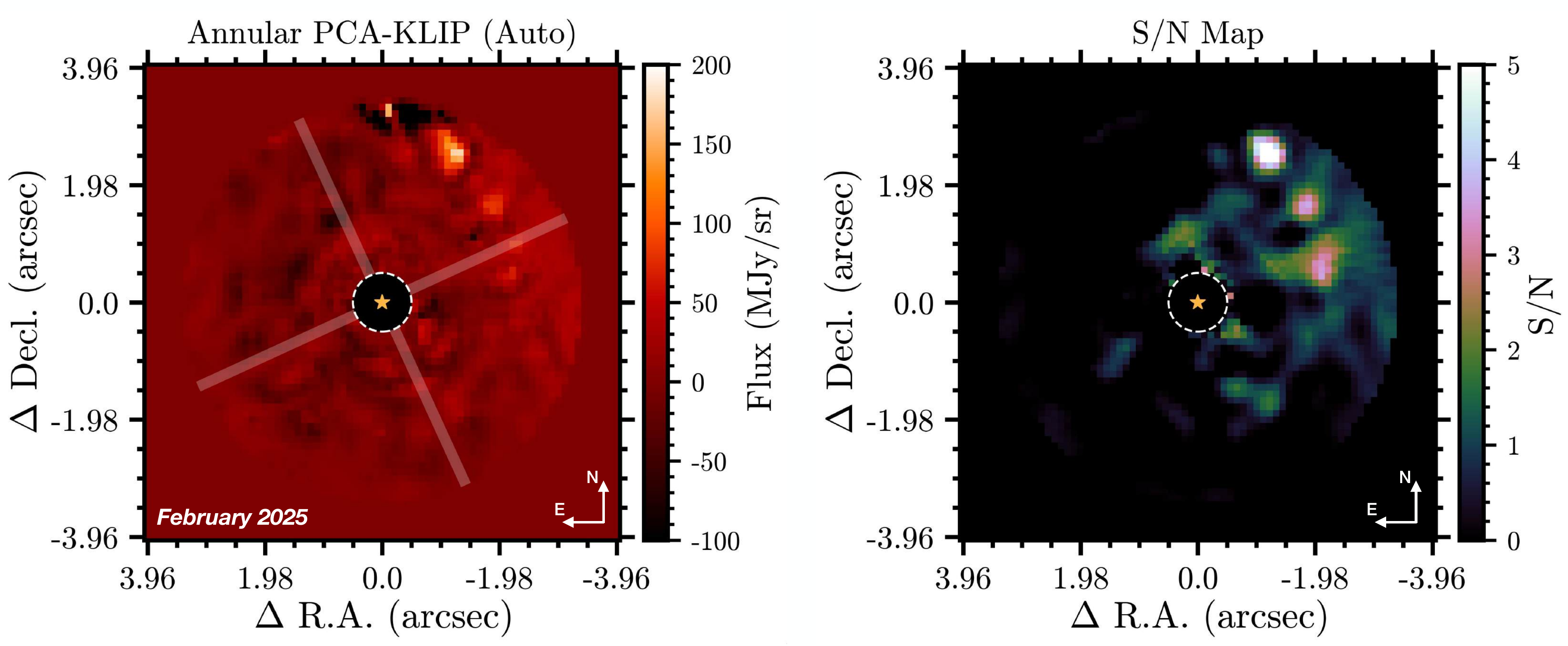}
\caption{Joint PCA-KLIP subtraction of \acenAB for the February 2025 observations with automatic principal component selection. \emph{Left:} The de-rotated residual image (North up, East left) after PSF subtraction in three concentric annuli. The MIRI/4QPM transition boundaries are shown as gray shaded regions. \emph{Right:} S/N map for the reduction shown in the left panel. No significant point sources are detected.}
\label{fig:joint-pca-feb}
\end{figure*}

\subsubsection{Detection of S1 in Multiple Data Subsets}
\label{sec:subset}
To verify that \objS is not a detector-level artifact in the \acenAB integrations, we perform two tests. First, we inspect the mean brightness in a 3-pixel radius aperture centered on \objS's position in the original, background-subtracted Stage 2 integrations. Sharp increases in the mean pixel brightness (from uncorrected hot pixels, for example) could result in a positive flux artifact at the location of \objS after PSF subtraction. The pixel brightness time-series shows that there are no such sharp transitions across the \acenAB integrations (Figure~\ref{fig:detector}). Second, we divide the (1250) \acenAB integrations into various subsets: first and second half of the data, even and odd frames, and first and last frame. Each subset is reduced independently using joint PCA-KLIP applied to a 3~FWHM width annulus centered at a radial separation of $\approx$1\farcs5 divided into two subsections along one of the 4QPM transition boundaries and using 200 PCs. A detector artifact would appear in some reductions, but not all. Performing this test, we detect \objS at a consistent S/N across all subsets (Figure~\ref{fig:subset-detections}). Together, the above two tests show that \objS is not likely a detector-level artifact in the \acenAB integrations. 

Interestingly, the above analysis shows that \objS can be detected in a single frame of the \acenAB integration sequence (exposure time of $\approx$7.5 seconds) with similar S/N. This indicates that the observations are not background-limited for the above exposure time. Additionally, the detection of \objS at the same position in the first and last integration of the $\approx2.6$ hour sequence \added{shows that that \objS is unlikely to be a foreground (Solar System) object such as a main-belt asteroid, which at the solar elongation of our observations ($\sim 100^\circ$) would have a mean proper motion of $\gtrsim$10 arcsec/hr\footnote{ \url{https://irsa.ipac.caltech.edu/data/SPITZER/docs/files/spitzer/asteroid_memo.pdf}}. Two other arguments suggest \objS is not a Solar System object. At a 12~\mum\ brightness level of $\geq$~3~mJy ($S$1 $\approx$~3.5 mJy at 15.5~\mum, see \S\ref{sec:measure}), there are fewer than $10^{-4}$ main belt asteroids in a 5\arcmin~$\times$~5\arcmin\ field at \acenA's ecliptic latitude of $\beta= -42^\circ$$^{15}$. Furthermore, the Minor Planet Catalog shows no known objects at the position of \acenA at the August epoch\footnote{\url{https://minorplanetcenter.net/cgi-bin/checkmp.cgi}}.}

\subsubsection{``Leave-One-Out" Analysis}
\label{sec:leave-out}
To verify that \objS is not an artifact introduced by any of the \emus coronagraphic reference images, we perform nine reductions where frames at each dither position of the \emus observation are iteratively excluded from the reference library. Each reduction is inspected to ensure that \objS is consistently detected in data. An artifact from the \emus coronagraphic images would be detected in some reductions, but not all. Performing the test with an identical reduction setup as in \S\ref{sec:subset}, we find that \objS is detected at consistent S/N across all reductions (Figure~\ref{fig:jackknife}) showing that it is not likely an \emus coronagraphic image artifact. A complete summary of all tests is provided in the conclusions (\S\ref{sec:conclusion}).

\begin{figure*}[htb!]
\centering
\includegraphics[width=\textwidth]{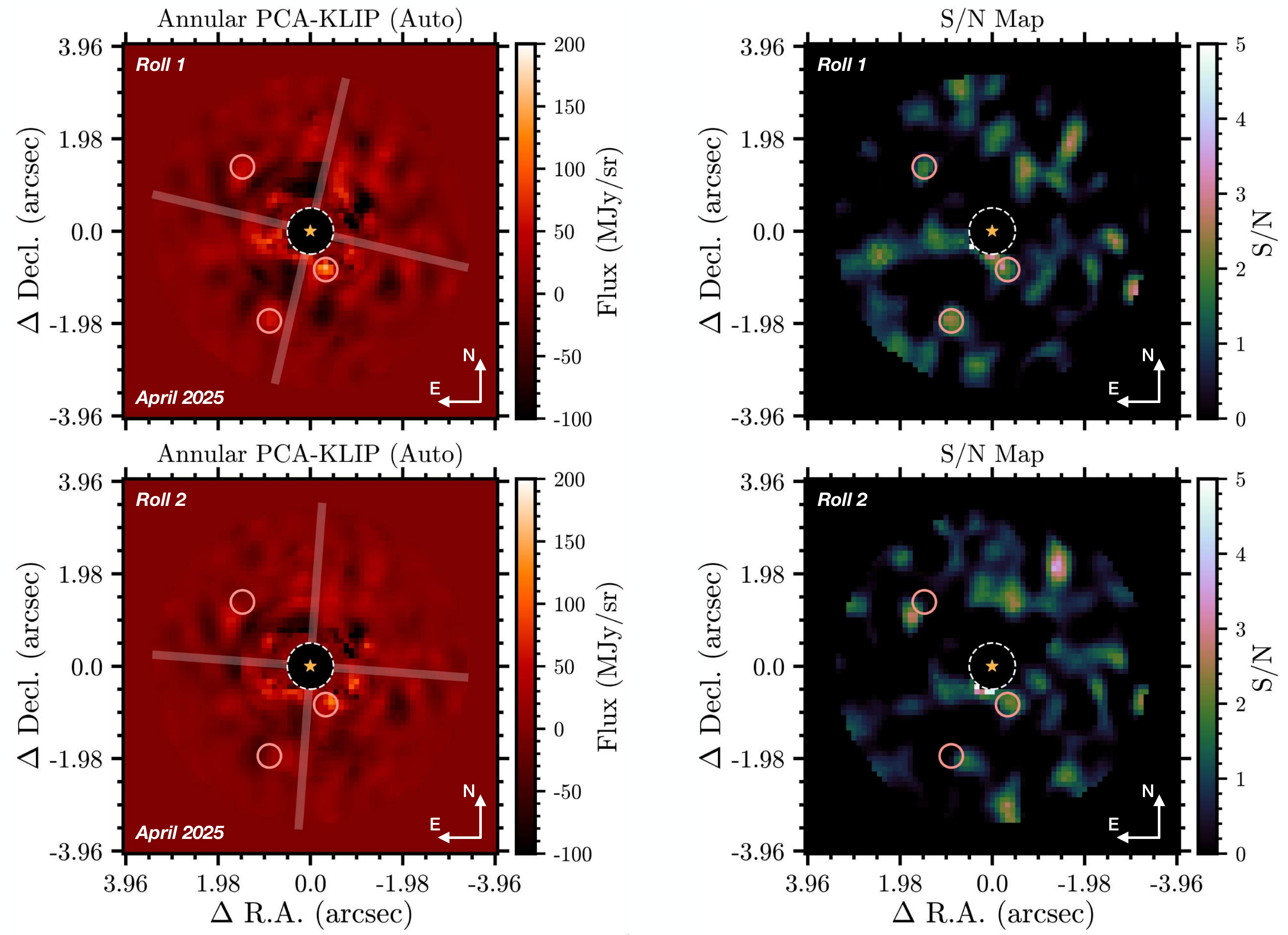}
\caption{Joint PCA-KLIP subtraction of \acenAB for the April 2025 observations with automatic principal component selection. \emph{Left column:} The de-rotated residual image (North up, East left) after PSF subtraction in three concentric annuli for each roll observation. The MIRI/4QPM transition boundaries are shown as gray shaded regions. Circular apertures are drawn centered on a few visually distinct features in the roll 1 reduction and at the same location in the roll 2 reduction. The features are present in the roll 2 reduction but offset in position angle (matching the direction of V3PA angle change between the two rolls) from their roll 1 location indicating that they are PSF subtraction artifacts. \emph{Right column:} S/N map for the reductions shown in the left column. No significant point sources are detected.}
\label{fig:joint-pca-apr}
\end{figure*}

\section{February and April 2025 Data Post-processing}
\label{sec:feb25-proc}
There were three primary goals for the February and April 2025 observations. (1)~Take advantage of \acenA's large proper motion ($\mu_\alpha,\; \mu_\delta=-3640,\;700$~mas~yr$^{-1}$) to test if \objS is a stationary background object; (2)~re-detect and confirm that \objS is an astrophysical source (orbiting planet) and not a residual image artifact; and (3)~conduct a new search for additional planets around \acenA that may be in more favorable orbital locations for detection.

\subsection{Is S1 a Background Object?}
\label{sec:background-s1a}
Using the ephemeris for \acenA (Beichman \& Sanghi et al. 2025, in press), we computed the location of \objS with respect to \acenA at the February and April 2025 epochs assuming it is a stationary background source (Figure~\ref{fig:background}). We then perform joint PCA-KLIP in an annulus of width 3 FWHM, centered at the expected radial separation for \objS as a background object, with the number of PCs automatically chosen as described in \S\ref{sec:auto-pc}. No point source is detected at the expected location of a stationary background source in either epoch (Figure~\ref{fig:background}). We confirm the sensitivity of our reductions by injecting a synthetic \webbpsf companion model with a flux of 2.5 mJy (1$\sigma$ lower bound for \objS, see Table~\ref{tab:fm}) in the \acenAB science cube, performing PSF subtraction identically as above, and unambiguously recovering the injected source (Figure ~\ref{fig:background}). This is definitive confirmation that \objS is not a background source.

\begin{figure*}[htb!]
\centering
\includegraphics[width=\textwidth]{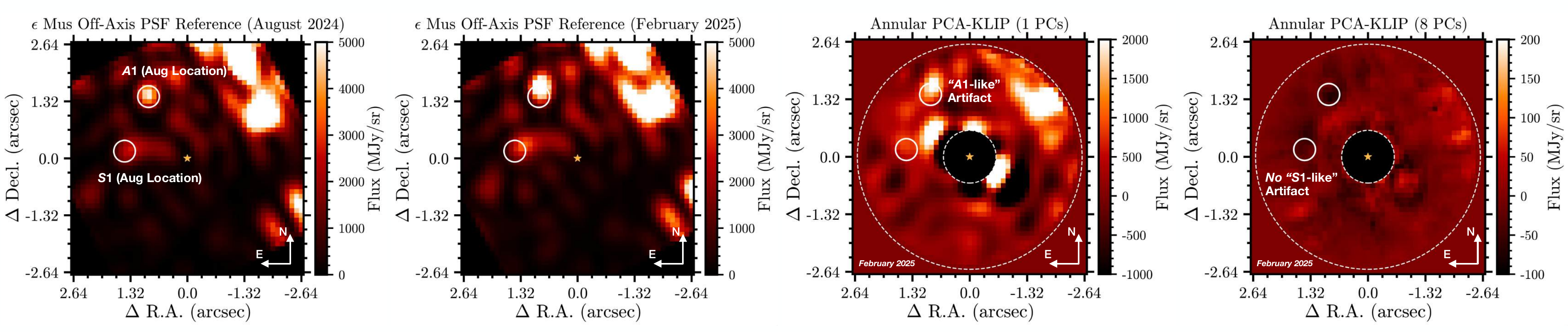}
\caption{\emph{Left:} the de-rotated off-axis \emus reference for \acenB for the August 2024 observations. It acts as a proxy for and shows the diffraction features expected from \acenB in vicinity of \acenA. The locations of detections \objS and \emph{A}1 in August 2024 are marked. \emph{Center left:} the de-rotated off-axis \emus reference for \acenB for the February 2025 observations. The diffraction pattern is nearly identical to that observed in August 2024. \emph{Center right:} joint PCA-KLIP reduction of the February 2025 dataset with 1~PC showing the presence of an ``\emph{A}1-like" artifact. \emph{Right:} joint PCA-KLIP reduction with 8 PCs showing that no persistent artifacts are seen at the location of \objS (and \emph{A}1).}
\label{fig:feb_b_psf}
\end{figure*}

\subsection{Is S1 Recovered as an Orbiting Planet?}
\label{sec:other-epochs}
To conduct a comprehensive search for point sources around \acenA, we perform reductions where joint PCA-KLIP is applied to a single annulus with a radial size of 2 FWHM ($\approx1\arcsec$) and centered at varying radial separations between 0\farcs75--3\farcs5 in steps of 0\farcs25 (twelve different reductions). We allow the \texttt{pca\_annulus} function in \texttt{vip\_hci} to auto-select the optimal number of principal components for each annulus as the value at which the standard deviation of the residuals (after the subtraction of the PCA approximation) drops below a tolerance value = 0.1. A S/N map is generated for each reduction following the definition in \citet{mawet_fundamental_2014} using the \texttt{snrmap} function in \texttt{vip\_hci}. For the April dataset specifically, each roll is reduced independently of the other using both \emus on-axis dithered observations and only the corresponding \emus off-axis observation. The de-rotated post-processed image for both rolls are then averaged to obtain the final reduction\footnote{We also tested angular differential imaging with the two roll April 2025 dataset. However, the mismatch between \acenA's position behind the mask between the two rolls (see \S\ref{sec:mask-pos}) results in significant residual artifacts, particularly along the 4QPM transition boundaries.}. 

No significant point sources were detected in our reductions of both the February and April 2025 datasets. We do not recover \objS as an orbiting planet in either follow-up epochs. As a summary, we show representative post-processed images (and the corresponding S/N maps) obtained by combining the reductions conducted in non-overlapping annuli centered at $\left\{1\arcsec, 2\arcsec, 3\arcsec\right\}$ (Figures \ref{fig:joint-pca-feb} and \ref{fig:joint-pca-apr}). A comparison of the reductions for each of the roll observations in April shows that PSF subtraction artifacts can be identified by their rotation in the sky plane (North up, East left coordinates). True astrophysical sources would remain stationary in each roll's de-rotated image. Overall, we find that the February and April 2025 reductions are less affected by residual artifacts from \acenB, primarily due to the better matching \emus off-axis reference observation (\S\ref{sec:align}). \added{Note that the reduction of the February 2025 dataset has been improved over the one presented in \citet{sanghi_preliminary_2025} with more careful pre-processing.}

\subsection{\acenB Diffraction Features in February 2025}
\label{sec:feb_diff}
Coincidentally, the difference between the telescope V3PA angles for the February 2025 and August 2024 observations of \acenAB was 181.8$^\circ$. The \acenB off-axis PSF is symmetric with respect to a $180^\circ$ rotation. Thus, its diffraction pattern, as viewed in sky coordinates (North up, East left), will be nearly identical between the two observation dates. Indeed, the two \acenB diffraction features co-located with \objS and \emph{A}1 in the August 2024 observations are also present, at approximately the same location, in the February 2025 \acenAB observations (Figure~\ref{fig:feb_b_psf}). If \objS is a persistent PSF artifact from \acenB in August 2024, it is reasonable to expect a similar artifact would be created after PSF subtraction in the February 2025 observations. PSF subtraction with the February 2025 observations shows that there is no persistent ``\objS-like" artifact at the location of the \acenB PSF diffraction feature (Figure~\ref{fig:feb_b_psf}). An ``\emph{A}1-like" artifact is created at very low PCs but is quickly subtracted out as the number of PCs increases (Figure~\ref{fig:feb_b_psf}). One difference between the August 2024 and February 2025 observations that could impact the interpretation of this analysis is that the February \emus off-axis reference required a smaller shift to align it with the \acenB PSF than the corresponding August off-axis reference (see \S\ref{sec:align}). Nevertheless, the above may be an independent argument against \objS being a residual PSF artifact in the August 2024 observations. 

\begin{figure*}[htb!]
\centering
\includegraphics[width=\textwidth]{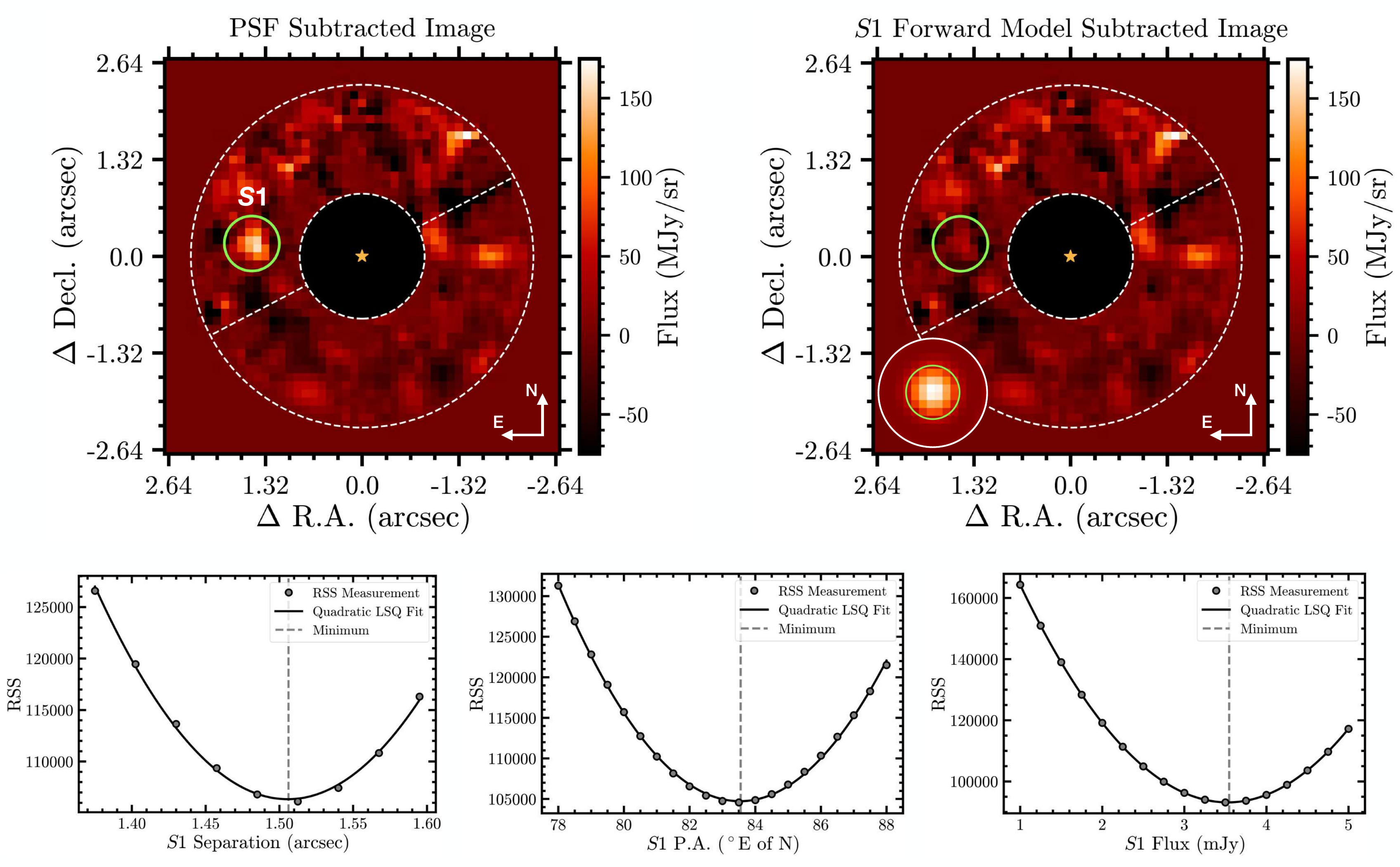}
\caption{\emph{Top panel:} the PSF subtracted image obtained before (left) and after (right) the injection of a negative copy of the best-fit \webbpsf forward model for \objS. The reduction parameters are identical between the two images. The \webbpsf model injected into the background-subtracted Stage 2 \acenAB integrations is pictured in the bottom left corner of the right image. The 1.5 FWHM diameter circular aperture where the residual sum of squares (RSS) is minimized is drawn for reference. \emph{Bottom panel:} marginalized RSS distributions as a function of injected model separation (left), P.A. (center), and flux (corrected for coronagraph throughput, right). The RSS distributions are well-fit by quadratic functions and the minimum corresponds to the forward modeled parameters.}
\label{fig:fm}
\end{figure*}

\section{Astrometry and Photometry of S1}
\label{sec:measure}
Astrometric and photometric measurements of \objS are extracted using the negative fake companion injection (NEGFC) method \citep[][]{lagrange_giant_2010, marois_exoplanet_2010}. This technique calibrates measurement biases introduced in the subtraction procedures. We apply this method to the MJy/sr calibrated images and use the joint PCA-KLIP reduction setup in \S\ref{sec:subset}.

First, we simulate a grid of PSF models for the \emph{S}1 candidate accounting for its position with respect to the MIRI/4QPM center using \webbpsf (with the appropriate MIRI filter and mask configuration, using the closest-in-time on-sky OPD map to the observations, and including detector effects). The PSF models are generated at radial separations ($\rho$) between 12.5--14.5 pixels (0\farcs11 per pixel) in steps of 0.5 pixels and position angles (P.A., $\theta$, measured East of North) between 78$^\circ$--88$^\circ$ in steps of 0.5$^\circ$. The \webbpsf models are normalized to the exit pupil of the optical system so that the sum of the PSF is 1.0 in an infinite aperture. The source throughput behind the MIRI coronagraph in F1550C at each position is estimated using \webbpsf simulations. We compute the ratio of integrated flux for a model PSF generated with and without the 4QPM inserted in the optical path at the given ($\rho$, $\theta$) position transformed to the detector frame.

\centerwidetable
\begin{deluxetable*}{cccccc}
\label{tab:fm}
\centering
\tablecaption{Astrometry and Photometry of \emph{S}1}
\tablehead{\colhead{$\rho$} & \colhead{$\theta$} & \colhead{$\Delta$R.A.\tablenotemark{\scriptsize a}} & \colhead{$\Delta$Decl.\tablenotemark{\scriptsize a}} & \colhead{$T_{\rm F1550C}$} & \colhead{$f_{\rm F1550C}$} \\ \colhead{(arcsec)} & \colhead{($^\circ$E of N)} & \colhead{(arcsec)} & \colhead{(arcsec)} & \colhead{} & \colhead{(mJy)}}

\startdata
$1.51\pm 0.13$ & $83.5\pm 4.9$ & $1.50\pm 0.13$ & $0.17\pm 0.13$ & 0.75 & $3.5\pm 1.0$ \\
\enddata

\tablenotetext{a}{$\Delta$R.A. and $\Delta$Decl. are analytically computed from the estimated ($\rho$, $\theta$) value. Positive $\Delta$R.A. is East and positive $\Delta$Decl.~is North.}

\end{deluxetable*}

Next, we iteratively inject negative copies of the simulated \webbpsf (with fluxes $f$ ranging from 1--5 mJy in steps of 0.25 mJy, accounting for coronagraph throughput\footnote{The flux in MJy/sr is corrected for coronagraph throughput $T$ and converted to mJy as: $f$~[mJy] $=$ $f$~[MJy/sr] $\cdot\;T^{-1}$ $\cdot$ \texttt{PIXAR\_SR}~[sr/pix] $\cdot\;10^9$, where \texttt{PIXAR\_SR} $= 2.86 \times 10^{-13}$ sr/pix.}) at each grid point into the background-subtracted \acenAB integrations, perform joint PCA-KLIP subtraction with identical parameters as in \S\ref{sec:subset}, and calculate the residual sum of squares (RSS) of pixel values in a fixed 1.5 FWHM diameter aperture, centered on the visually approximated centroid of \objS. Once we have obtained the RSS for each ($\rho$, $\theta$, $f$) grid point, we marginalize the 3D distribution of RSS values across each axis to obtain the distribution of RSS as a function of separation, P.A., and flux (Figure~\ref{fig:fm}). The RSS follows a parabolic function of the injected separation, P.A., and flux which is the expected outcome when the measurement biases are caused by over-subtraction \citep{apai_high-cadence_2016, pueyo_detection_2016, zhou_hubble_2021, sanghi_efficiently_2022}. We perform a quadratic least-squares (LSQ) fit to the measurements and use the minimum of each best-fit parabola as the final measurements for the separation, P.A., and flux density of \objS (Table~\ref{tab:fm}). The best-fit \webbpsf model cleanly subtracts \objS's signal as seen in Figure~\ref{fig:fm}. This shows that \objS is well-represented as a point source.

The photometric uncertainty consists of four components: the speckle noise ($\sigma_{sp}$), the photon noise ($\sigma_p$), the read noise ($\sigma_r$), and a PSF normalization uncertainty ($\sigma_n$). All uncertainties are quoted as a fraction (in \%) of the point source flux. The speckle noise is estimated as the standard deviation of integrated flux in a series of non-overlapping 1.5 FWHM diameter apertures (same size as used for flux estimation of \objS) constructed at the radial separation of \objS in the final joint PCA-KLIP PSF-subtracted image. We find that the speckle noise is $\approx$25\%. The photon and read noise are estimated across pixels in a 1.5 FWHM diameter aperture centered on \objS (in the detector frame) by appropriately combining the per-pixel variance provided in the \texttt{VAR\_POISSON} and \texttt{VAR\_RNOISE} extension images in the background-subtracted Stage 2 data product. These two terms are negligible ($>$50$\times$ smaller than the speckle noise). Finally, the PSF normalization uncertainty is associated with conversion from detector units DN/s to physical units MJy/sr. We adopt an uncertainty of 11\% based on the results in \citet{malin_first_2024}. We assume the above sources of uncertainty are independent and take the final uncertainty on the photometry as the quadratic sum of the two dominant terms: $\sigma_f = \sqrt{\sigma_{sp}^2 + \sigma_n^2}$. For astrometric uncertainty estimation, we assume the radial and tangential directions are independent and calculate their uncertainties as FWHM/SNR $= 0\farcs5/4 \approx 0\farcs13$. We note that the error in stellar position behind the MIRI/4QPM is $\sim$10~mas (\S\ref{sec:mask-pos}), which is negligible compared to the above uncertainty. The final results are presented in Table~\ref{tab:fm}. \added{Additionally, we note that \objS's astrometry and photometry was independently verified (within 1$\sigma$) using the \texttt{pyKLIP} package \citep{wang_pyklip_2015, pueyo_detection_2016, wang_orbit_2016} and confirmed to be consistent between the joint PCA-KLIP reduction and the classical median-RDI reduction.}

\begin{figure*}[htb!]
    \centering
    \includegraphics[width=\textwidth]{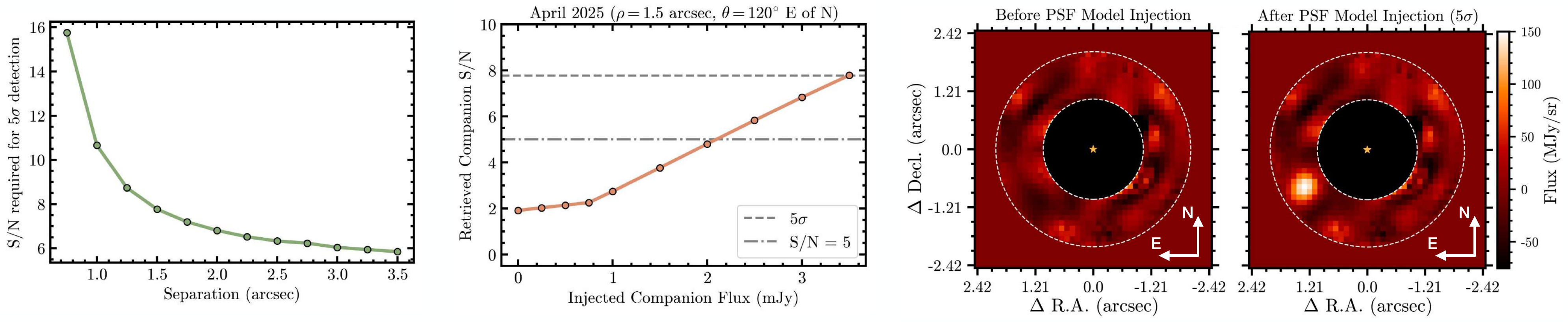}
    \caption{Illustrative example of the injection-recovery procedure. \emph{Left:} the S/N corresponding to a 5$\sigma$ detection for the radial separations at which synthetic companions are injected. \emph{Center:} S/N of a synthetic companion injected at a given (example) position as a function of flux. The 5$\sigma$ and S/N = 5 detection thresholds are marked with horizontal dashed and dash-dotted lines, respectively. \emph{Right:} PSF-subtracted images before and after the injection of the model PSF corresponding to a 5$\sigma$ detection ($\approx$3.5 mJy in this case).}
    \label{fig:contrast-demo}
\end{figure*}

\section{Point Source Sensitivity Analysis}
\label{sec:sensitivity}
\subsection{Injection-Recovery Procedure}
We perform PSF injection-recovery tests with \webbpsf models to characterize the flux/contrast sensitivity of our observations. PSF injections are carried out over the following grid, defined in polar coordinates: $\rho \in$ [0\farcs75, 3\farcs5] in steps of 0\farcs25 and $\theta \in$ [0$^\circ$, 350$^\circ$] in steps of 10$^\circ$. We report our sensitivity as (a) the point source flux detectable at a S/N = 5; and (b) the point source flux detectable at a 5$\sigma$ significance level. We use \texttt{vip\_hci}'s \texttt{significance} function to calculate the S/N (accounting for small sample statistics) that corresponds to a 5$\sigma$ Gaussian detection significance (for the equivalent false positive probability, $\sim3\times10^{-7}$) at each radial separation of injection (Figure~\ref{fig:contrast-demo}). The S/N = 5 and 5$\sigma$ flux sensitivity is converted to contrast sensitivity by dividing by \acenA's flux in the F1550C filter ($f_* = 63$~Jy; Beichman \& Sanghi et al. 2025, in press). To determine the sensitivity at a given ($\rho$, $\theta$), the general procedure for all three datasets is as follows:
\begin{enumerate}
    \item Inject a \webbpsf companion model generated for the appropriate position on the MIRI detector, filter and mask configuration, using the closest-in-time on-sky OPD map, including detector effects, and scaled to a flux $f$ in mJy (after correcting for throughput behind the 4QPM coronagraph at the injection location).

    \item Perform a joint PCA-KLIP reduction in an annulus of width 2~FWHM centered at the injection separation with the number of PCs automatically chosen, as discussed in \S\ref{sec:auto-pc}. For the April observations, each roll is reduced independently and the PSF subtracted images are averaged to obtain the final detection image.

    \item Estimate the S/N at the location of injection, accounting for small sample statistics.

    \item Repeat step 3 for multiple values of $f$ and identify the companion flux for which the target S/N computed earlier is achieved (Figure~\ref{fig:contrast-demo}).
\end{enumerate}

\begin{figure*}[htb!]
\centering
\includegraphics[width=\textwidth]{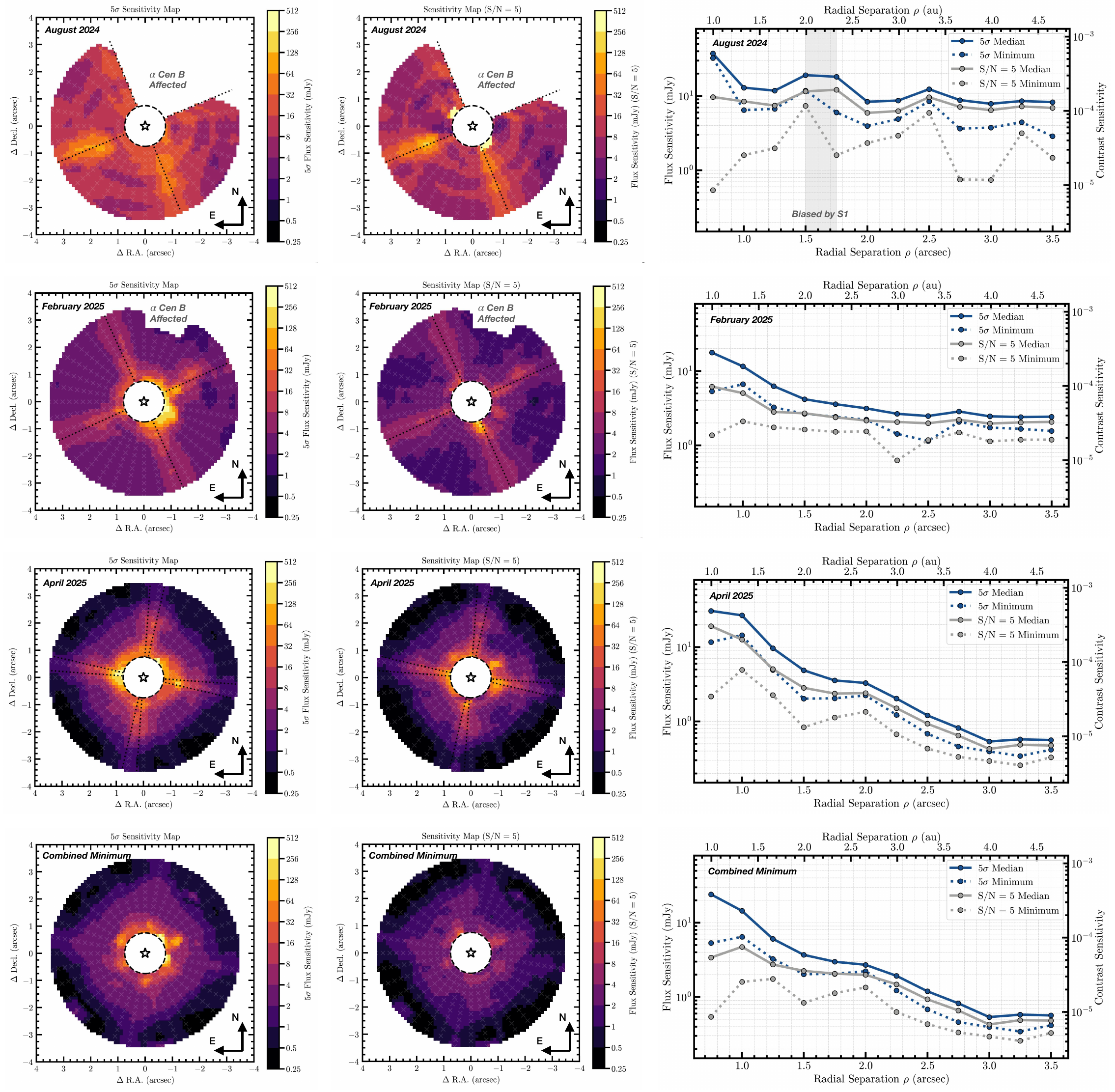}
\caption{Results of injection-recovery analysis with the \acenAB data. \emph{Left column:} two-dimensional 5$\sigma$ flux sensitivity maps for each of the three epochs of observation, and their combined minimum, in sky coordinates (North up, East left). The central region ($<$ 0\farcs75, or $<$ 1.5 FWHM, radial separations) is masked (poor detectability). The dotted lines correspond to the orientation of the 4QPM transition boundaries (one for each roll in April 2025). The regions excluded from the injection-recovery analysis in the August 2024 and February 2025 observations due to contamination from \acenB are labeled. The crosses mark the ($\rho$, $\theta$) grid points where sensitivity was estimated. A discrete colormap is chosen to highlight the different sensitivity zones across the image. \emph{Center column:} same as the left column for the S/N = 5 flux sensitivity. \emph{Right column:} 5$\sigma$ and S/N = 5 median and minimum flux/contrast curves for each epoch. \added{The median and minimum flux/contrast are both computed over all position angles of injection for a given separation from the 2D sensitivity maps.}}
\label{fig:sensitivity}
\end{figure*}

We note here modifications made to the above procedure for the August 2024 and February 2025 observations, which were impacted by contamination from \acenB's PSF in the vicinity of \acenA. The \emus off-axis reference in the August 2024 observation required an $\approx$3 pixel shift to align it with \acenB (\S\ref{sec:align}). This results in an imperfect subtraction of \acenB (specifically, for the brightest diffraction features) and generates several artifacts in the North-West quadrant of the image (see Figure~\ref{fig:joint-pca}, for example), which impact the detectability of point sources in the region and bias the S/N calculations for injected sources at other position angles. Thus, we exclude this area (P.A.~$= 290^\circ$--$20^\circ$) from the injection-recovery tests for all injection separations. Similarly, the February 2025 reductions are affected by a residual subtraction artifact from \acenB at P.A.s~between $330^\circ$--$0^\circ$ (see Figure~\ref{fig:joint-pca-feb}, for example) for separations $\geq$~2\farcs75. This area is excluded from the injection-recovery tests. In both cases, the noise estimation for S/N calculations of the injected sources at other position angles is conducted using apertures that do not fall in the excluded region. In principle, we could manually tune (increase) the number of PCs to better subtract the residual \acenB artifacts (as can be seen from comparing the low and high PC reductions in Figure~\ref{fig:s1-a1}; \added{see also discussion in Appendix~\ref{sec:app-bin}}). However, this is impractical to implement given the number of injection-recovery tests being conducted.

\subsection{2D Sensitivity Maps}
From the above analysis, we found that the sensitivity to companions in the vicinity of \acenA can vary significantly as a function of spatial position due to the complex nature of the underlying speckle field (e.g., because of the absence of a second roll angle observation and/or the contaminating presence of \acenB) and the effect of the 4QPM transition boundaries. Thus, we report our S/N = 5 and 5$\sigma$ sensitivities in the form of two-dimensional (2D) maps (Figure \ref{fig:sensitivity}). For a few injection locations in each of the epoch observations, we were unable to accurately determine a sensitivity. Three reasons were (1)~the S/N threshold for 5$\sigma$ detection was not met, even for low contrast ($\sim 10^{-1}$) injections (primarily at 0\farcs75 separation, which requires S/N~$\approx$~16 for a 5$\sigma$ detection, Figure~\ref{fig:contrast-demo}); (2)~the presence of a known PSF subtraction artifact at the injection location positively biased the flux and led to a significantly deeper S/N~=~5 flux sensitivity estimate than appropriate; (3) in the case of the August 2024 observations, injections near the location of \objS were biased. All of the above cases were identified by inspecting the PSF-subtracted images for the different flux levels of injection at each position. We do not report sensitivities at these locations. The final 2D sensitivity maps are constructed by transforming the polar coordinate grid to a rectangular grid and linearly interpolating sensitivity values (using \texttt{scipy.interpolate.griddata}) at a resolution of 0\farcs11/pixel (equal to the MIRI detector platescale). While it is most accurate to use the 2D map for sensitivity analyses, we also present traditional 5$\sigma$ (and S/N = 5) median flux/contrast curves (Figure \ref{fig:sensitivity}) as they are commonly used in literature. We compute the median (azimuthally averaged) contrast as a function of separation. We also present the azimuthal minimum sensitivity as a function of radial separation.

The 2D sensitivity maps (Figure \ref{fig:sensitivity}) show remarkable variation between the various observation epochs. The August 2024 maps are non-uniform, with the sensitivity achieved being highly dependent on the exact location in the image. This is a consequence of the higher contamination from \acenB due to its imperfect subtraction. The February 2025 maps are more uniform and demonstrate a better sensitivity to point sources compared to August 2024 because of the improved subtraction of \acenB. The sensitivity is relatively flat at separations $>$2\arcsec (see the contrast curve). The April 2025 maps are uniform, radially striated, and achieve deeper sensitivity than the February 2025 observations at separations $>$2\arcsec. The biggest reason for this difference is the significantly lower contamination from \acenB due to its orientation in the detector plane (see Figure~\ref{fig:fullfr}). The availability of two roll angles of observation also contributes to the improved sensitivity. A comparison of the contrast curves shows that the February 2025 observations achieve better sensitivity than the April 2025 observations at separations $<1\farcs25$. At close-in separations, the latter epoch is compromised by residual artifacts from PSF subtraction (see Figure~\ref{fig:joint-pca-apr}). A common feature across all the maps is the poor sensitivity along the 4QPM transition boundaries, as expected, because of the low planet throughput. Our observations typically reach a 5$\sigma$ sensitivity better than 4 mJy corresponding to contrast ratios deeper than 6$\times 10^{-5}$ for separations $\gtrsim$1\farcs5. Our best sensitivities are $\lesssim$2 mJy corresponding to contrast ratios deeper than 3$\times 10^{-5}$ for separations $\gtrsim$1\farcs5. The above are comparable to, and in some cases better than, those achieved for single star F1550C observations with JWST/MIRI \citep[e.g.,][]{boccaletti_jwstmiri_2022, malin_first_2024, matthews_temperate_2024}. In general, at the above separations, F1550C observations are background-limited \citep{boccaletti_jwstmiri_2022}. The improvement in contrast sensitivity over past observations is because of the lower background limit given the bright stellar magnitude of \acenA. The 5$\sigma$ and S/N = 5 sensitivities differ by about a factor of 2 at separations $<1\farcs25$ and the difference is lower for larger separations. As a summary, we also present an epoch-combined sensitivity map (Figure \ref{fig:sensitivity}), showing the best sensitivity across epochs at a given injection location, and the corresponding contrast curves. \added{It is important to note that the sensitivities presented for all three epochs are conservative (particularly August 2024). Better sensitivity can be achieved by manually tuning (increasing) the number of principal components used to improve the suppression of subtraction artifacts generated by both \acenA and \acenB (as was done to test the robustness of the \objS candidate in \S\ref{sec:pcs-vary} and demonstrated in Appendix~\ref{sec:app-bin}). As noted earlier, this is not practically feasible for our injection-recovery tests.}

Overall, the above analysis highlights the importance of observing a system like \acenAB, where the sensitivity achieved is \emph{inconsistent} between different observations (due to the unique binary astrophysical scene), at multiple epochs to improve completeness to planets that may be located at unfavorable locations (e.g., near the transition boundary, within the inner working angle, or at the location of artifacts) in any single epoch. The individual epoch sensitivity maps are particularly important in the context of \acenAB as they constrain the orbits of candidate \objS consistent with non-detections in February 2025 and April 2025. All individual epoch (interpolated) 2D sensitivity maps presented in this paper, as well as the sensitivities at the individual ($\rho$, $\theta$) grid points where PSF injection-recovery tests were performed, are available at\dataset[10.5281/zenodo.15748931]{https://zenodo.org/records/15748931}. The orbital analysis, along with a translation of the achieved flux sensitivity to detectable planet scenarios (effective temperature and radius), is presented in Paper I (Beichman \& Sanghi et al.~2025, in press).

\begin{figure}[tb!]
\centering
\includegraphics[width=\linewidth]{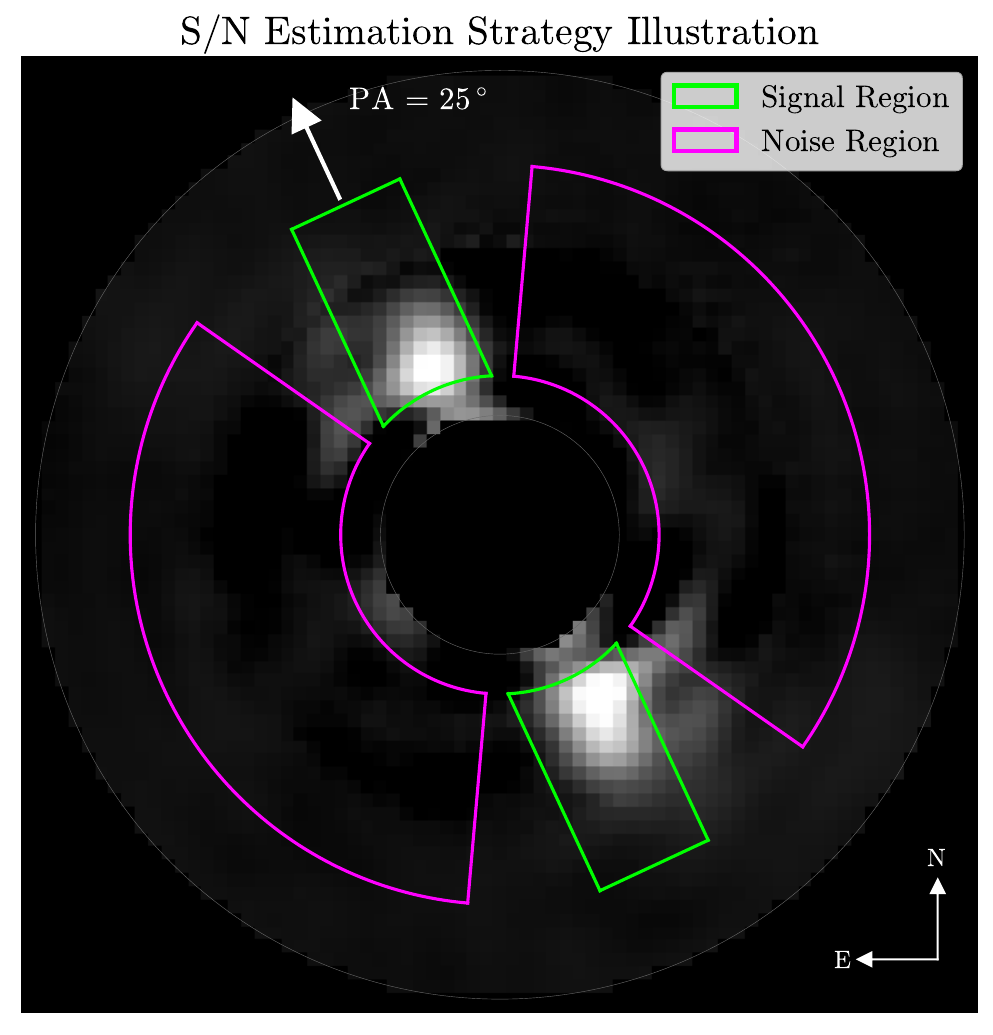}
\caption{Illustration of how the S/N map is generated for the detection of extended emission. The figure shows an example of a signal region (green) at one position angle, along with the corresponding perpendicular annular noise region (purple), using the coplanar ABA-1 exozodi model injected into the April data. The S/N is calculated by varying the aperture's position angle and length.}
\label{fig:ilust-snr}
\end{figure}

\section{Search for and Limits on Exozodiacal Emission}\label{sec:snr_exozodi}

\begin{table*}[!tb]
    \centering
    \caption{Signal-to-noise (S/N) values and empirical significance (in units of $\sigma$) of different exozodi models and geometries at each observation epoch, alongside zodi levels for the different models.}
     \begin{tabular}{lccccc}
    \hline
    \hline
     Geometry & ABA-1 & ABA-2 & ABA-3 & ABA-4 & ``1-zodi" \\
    \hline
    \multicolumn{6}{c}{February 2025 (7.4 = 2.1$\sigma$)\tablenotemark{\scriptsize a}} \\
    \hline
     Coplanar & 40.5 (11.6$\sigma$) & 27.3 (7.9$\sigma$)  & 10.0 (2.9$\sigma$) & 3.0 (0.9$\sigma$) & 0.4 (0.1$\sigma$) \\
     Offset   & 73.6 (21.2$\sigma$) & 56.0 (16.1$\sigma$) & 27.6 (7.9$\sigma$) & 10.2 (2.9$\sigma$) & 4.3 (1.2$\sigma$) \\
    \hline
    \multicolumn{6}{c}{April 2025 (5.0 = 1.3$\sigma$)\tablenotemark{\scriptsize a}} \\
    \hline
     Coplanar & 78.7 (20.8$\sigma$) & 52.6 (13.9$\sigma$) & 21.3 (5.6$\sigma$) & 5.8 (1.5$\sigma$) & $-0.7$ ($-$0.2$\sigma$) \\
     Offset & 86.1 (22.8$\sigma$) & 46.8 (12.4$\sigma$) & 14.8 (3.9$\sigma$) & 4.4 (1.2$\sigma$) & 0.6 (0.2$\sigma$) \\
    \hline
    \hline
    Zodi Level &  &  &  &  &  \\
    \hline
    $Z_L$\tablenotemark{\scriptsize b} & 58 & 17 & 5.1 & 1.8 & 0.94 \\
    $Z_{\Sigma}$\tablenotemark{\scriptsize c} & 84 & 29 & 8.4 & 3.0 & 1.0  \\
    \hline
    \end{tabular}
     \tablenotetext{a}{Overall maximum S/N (and corresponding empirical significance) found in the original post-processed data, with no model disk injected.}
    \tablenotetext{b}{The luminosity-based zodi level $Z_L$ is defined as the ratio of the exozodi's fractional luminosity to that of the Solar System's zodiacal cloud.}
    \tablenotetext{c}{The surface density-based zodi level $Z_{\Sigma}$ is given by the ratio of the disk surface density at the Earth-equivalent insolation distance---approximately $1.23\,\text{au}$ for \acenA---to that of the Solar System at $1\,\text{au}$.}
\label{tab:sn_models_epochs}
\end{table*}

We investigate the detectability of faint, spatially extended exozodiacal emission in the \acenAB system in the February and April~2025 epochs. This section introduces our analysis framework, including an empirical approach to estimating detection significance, which we also use in injection–recovery tests to quantify sensitivity to varying  levels of exozodiacal emission.

\subsection{S/N Map Calculation}\label{sec:snr_map}

\added{ To evaluate the presence of a potential signal in both the original and model-injected data, we compute the signal-to-noise ratio (S/N) in a certain aperture across a grid of predefined aperture geometries, hereafter referred to as the S/N map. The target aperture geometry is motivated by the expected morphology of an inclined exozodiacal disk and consists of a star-centered rectangle, as illustrated in Figure~\ref{fig:ilust-snr}. By systematically varying the aperture's length and orientation, we compute the S/N map, as detailed in the steps below:

\begin{enumerate}
    \item We integrate the flux inside a given target aperture (signal region), which is defined as a rectangle with a width of 1\arcsec ($\approx 2$~FWHM) and an overall length of 3\farcs8--6\arcsec, with a central region of radius 1\farcs3 masked to avoid residual PSF subtraction artifacts. Converted to projected distance from \acenA, our aperture stretches roughly from 1.7~au at the inner boundary out to 2.4--4~au.
    
    \item We then place another aperture orthogonal to the target aperture to determine a background and noise level. This aperture has the geometry of two opposing annulus sectors, each with a radial extent equal to the target aperture handles, and an angular extent of 120 degree (Figure~\ref{fig:ilust-snr}).
    
    \item The S/N is computed using the formula $\mathrm{S/N = (signal - background) / noise}$
    
    \item We repeat steps 1--3, varying the position angle of the target aperture from 0$^\circ$ to 180$^\circ$, as well as the target aperture length from 3\farcs8--6\arcsec.
\end{enumerate}

For both the February and April~2025 observations, we mask a 0\farcs5 width corridor centered on the MIRI 4QPM transition boundaries when computing the S/N maps. Additionally, for the February~2025 observations, we mask the North-West region that is impacted by residuals from \acenB's PSF. Note that we are not using any prior knowledge about any potential (or injected) disk properties (position angle or size) in our calculations.}

\begin{figure*}[htb!]
\centering
\includegraphics[width=\textwidth]{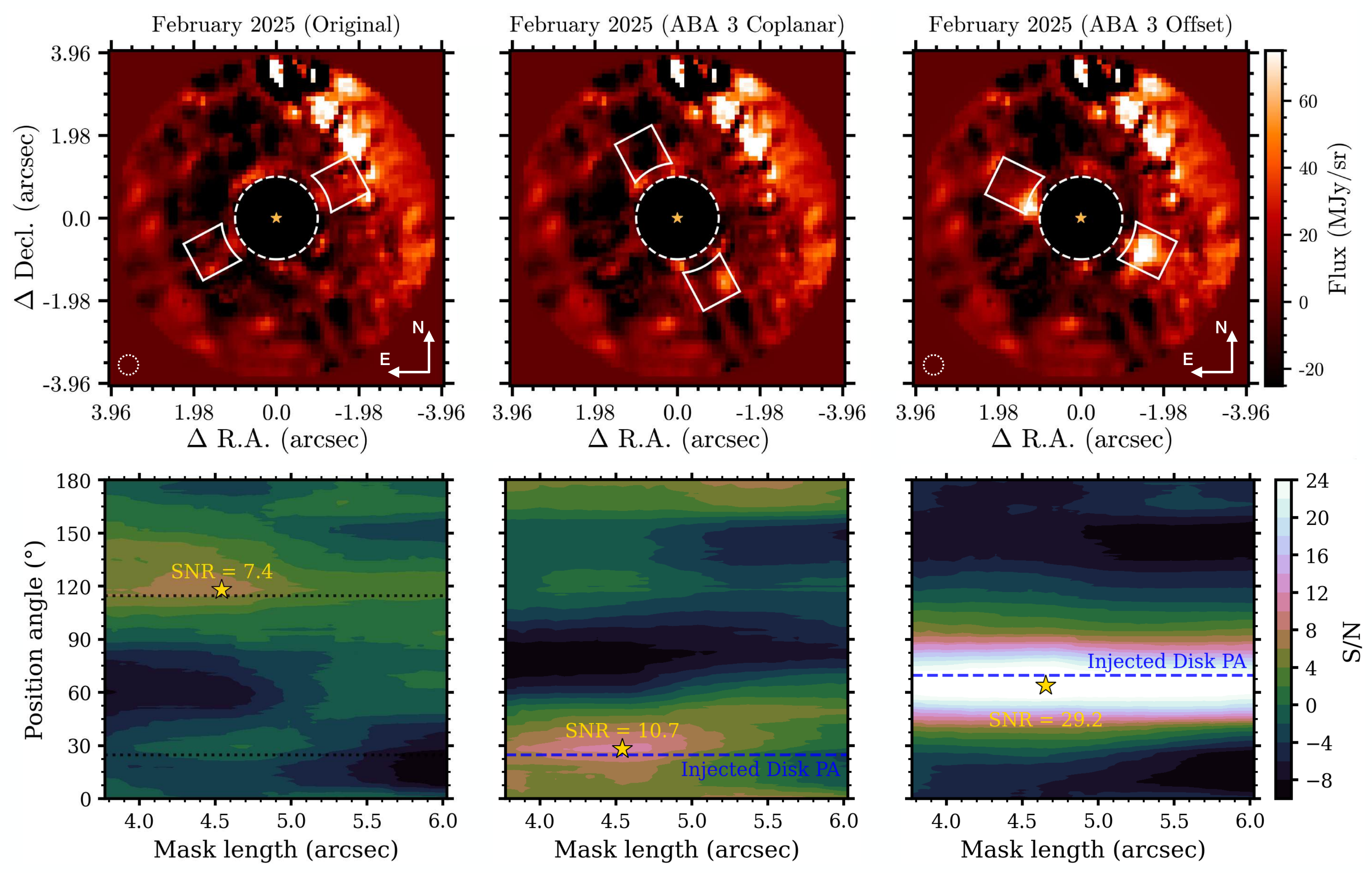}
\caption{PSF-subtracted images (top) and the corresponding extended emission S/N maps (bottom) for the February 2025 observation for three cases: (left) original data, (center) ABA-3 exozodi model injected with coplanar orientation, and (right) ABA-3 exozodi model injected with offset orientation. In the image plots, a 1~FWHM diameter beam is indicated in the bottom left corner with a dotted circle. The aperture that yields the highest S/N (marked by a star in the S/N maps) is shown in the images. For the original data, dotted lines in the S/N map mark the position angles of the 4QPM transition boundaries. For the injected cases, dashed lines mark the position angles of the injected disks.}
\label{fig:snr_feb}
\end{figure*}

\begin{figure*}[htb!]
\centering
\includegraphics[width=\textwidth]{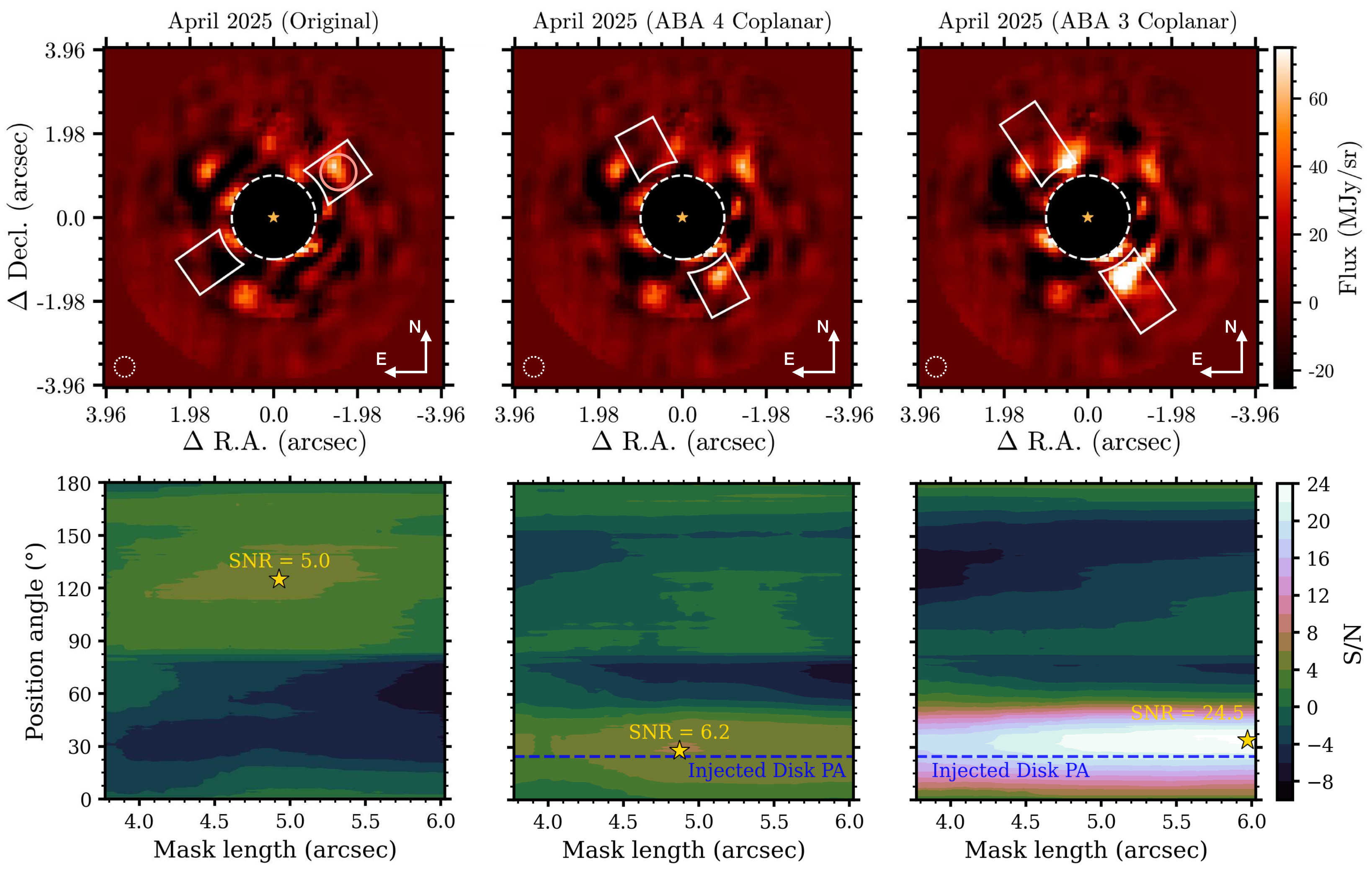}
\caption{Same as Figure~\ref{fig:snr_feb} for the April 2025 observations. Here, the injected cases shown are the ABA-4 and ABA-3 model, both in the coplanar orientation. In the image showing the original data (top left), a solid circle marks a suspected PSF-subtraction artifact (see the text).}
\label{fig:snr_apr}
\end{figure*}

\subsection{S/N of Extended Emission in the Original Data}\label{sec:snr_orig}

The February and April 2025 epoch observations are the two best datasets to search for extended emission given the better matching reference for and subtraction of \acenB (see \S\ref{sec:align}, \S\ref{sec:other-epochs}, and Figure~\ref{fig:sensitivity}). Unlike a close-in orbiting planet, the properties of extended emission features are not expected to change between observations on a one-year timescale. Hence, we focus our reduction efforts on the February and April 2025 epoch observations. For each epoch, we perform a single reduction where joint PCA-KLIP is applied in two concentric annuli each with a radial size of 3~FWHM ($\approx1.5\arcsec$), centered at radial separations $\left\{1\farcs75, 3\farcs25\right\}$, and with the number of PCs automatically selected as described in \S\ref{sec:auto-pc}. The central 2~FWHM radius region is masked. For the April dataset, each roll is reduced independently and then the post-processed images are stacked. A S/N map across aperture geometries is generated for each epoch following the method described in the previous section. The results are shown in the left columns of Figures~\ref{fig:snr_feb} and~\ref{fig:snr_apr}. 

\added{The peak S/N is 7.4 for the February observations and 5.0 for the April observations. To assess detection significance, we define an empirical signal-to-noise significance, $\sigma$, by normalizing each peak S/N value by the 68.3 percentile of the absolute S/N distribution in the respective PSF-subtracted images. This threshold represents the empirical 1$\sigma$ level of background fluctuations under the full measurement process, including effects of correlated noise, and is $\text{S/N}_{1\sigma}\!=\!3.5$ and $\text{S/N}_{1\sigma}\!=\!3.8$ for the February and April~2025 data, respectively. We then define a detection as a significance threshold of $\sigma\!>\!3$, that is $\text{S/N}_{3\sigma}\approx\!11$ in both cases. It thus follows that no significant extended emission features are detected in either epoch. We note that the position angle along which the peak S/N is obtained in the February~2025 image coincides with one of the 4QPM transition boundaries. Additionally, the peak S/N in the April~2025 image is biased by a residual PSF subtraction artifact (Figure~\ref{fig:snr_apr}), rather than arising from disk emission, as was verified by the rotation of the feature between the reductions for each roll in the sky orientation (North up, East left). A true astrophysical disk would remain fixed in sky orientation between the two rolls.}

\subsection{S/N with Injected Disks}\label{sec:snr_inj}

We investigated the sensitivity of the February and April 2025 epoch observations by injecting five exozodi disk models at two geometries: four asteroid belt analogue models and a fiducial ``1-zodi" model, at ``coplanar'' and ``offset'' geometries, respectively. The input physics and model calculation procedures are described in Paper I (Beichman \& Sanghi et al. 2025, in press). \added{In summary, we first simulate surface brightness distributions for different disks, considering (1)~a set of asteroid belt analogue (ABA) models, in which dust is collisionally produced in a narrow belt and transported inward via Poynting–Robertson drag \citep[following][]{rigley_dust_2020}, and (2)~a Solar system analogue based on the \citet{kelsall_cobe_1998} zodiacal cloud model, placed around \acenA. For the ABA models, we vary the belt mass across four levels (ABA-1 to ABA-4), where ABA-1 corresponds to the most massive and brightest disk, and ABA-4 to the least massive and faintest disk. We then synthesize 4QPM coronagraphic observations of the inclined disks \citep[following][]{sommer_pr_2025}, for injection into our datasets}. Relevant to the analysis presented here, the coplanar geometry refers to an exozodi model with the same inclination and position angle as the \acenAB binary. The offset geometry refers to an exozodi model with the same inclination as the \acenAB binary but offset in position angle by 45$^\circ$, counter-clockwise. We note here that the coplanar orientation falls along the 4QPM transition boundary for the February 2025 observations (but not the April 2025 observations). The offset geometry models were generated to quantify the sensitivity level in regions not impacted by throughput-loss due to the transition boundaries for the February 2025 epoch. 

We iteratively injected each exozodi model, at each geometry ($5\times 2$), into the original \acenAB integrations (for each epoch), performed the same joint PCA-KLIP reduction described in the previous section, and again computed the S/N map across search aperture geometries, \added{as well as significances with respect to the $\text{S/N}_{1\sigma}$ from the original data, as described in \S\ref{sec:snr_map} and \S\ref{sec:snr_orig}. Table~\ref{tab:sn_models_epochs} summarizes the retrieved S/N and empirical significance for each model injection case for both epochs. We find that the exozodi models ABA-3 and brighter are consistently above or near the detection threshold.
In the April~2025 observation, the ABA-4 model is not detected, whereas the ABA-3 model is readily detected (Figure~\ref{fig:snr_apr}). The $3\sigma$ threshold is narrowly missed in the February epoch for the coplanar disk. This is because, in that epoch, the 4QPM transition boundary aligns precisely with the disk and significantly attenuates its flux. This is illustrated in Figure~\ref{fig:snr_feb}, showing the injected images and S/N maps for the coplanar as well as the offset ABA-3 model.}

Overall, our injection-recovery tests place an upper limit \added{$Z_L \lesssim$~5~zodis and $Z_\Sigma \lesssim$~8~zodis (ABA-3 model, Table~\ref{tab:sn_models_epochs})} on the presence of (coplanar) exozodi around \acenA. This level of sensitivity is remarkable: $>$5--10$\times$ deeper than the current best limits achieved with nulling interferometry using the Large Binocular Telescope Interferometer \citep{ertel_hosts_2018, ertel_hosts_2020}. The primary reason for the above gain in sensitivity is JWST's ability to resolve \acenA's habitable zone (due to the star's proximity) and JWST's excellent stability, which allows us to accurately subtract the stellar contribution and achieve deep contrast performance. A detailed discussion of the implications of this upper limit is presented in Paper I (Beichman \& Sanghi et al. 2025, in press).

\section{Conclusions}
\label{sec:conclusion}

We conducted JWST/MIRI F1550C coronagraphic imaging observations of our closest solar twin, $\alpha$ Centauri A, over three epochs between August 2024 and April 2025 to directly resolve \acenA's habitable zone and perform a deep search for planets and exozodiacal disk emission. In this paper, for the first time with JWST, we demonstrated the application of reference star differential imaging to simultaneously subtract the coronagraphic image of a primary star (\acenA) and the PSF of its binary companion (\acenB). The post-processing strategies used in this work were specifically developed to tackle the uniquely challenging and dynamic binary astrophysical scene presented by \acenAB. The key results from our program are summarized below.

\textbf{Point source sensitivity.} We overcome extreme contamination from \acenB and achieve a typical 5$\sigma$ point source contrast sensitivity between $10^{-5}$ and $10^{-4}$ for separations $\gtrsim$~1\arcsec, not only comparable to, but in some cases better than, those achieved for single star F1550C observations with JWST/MIRI. The improvements in contrast sensitivity are unique to \acenA given its bright stellar magnitude, which yields a lower background limit in contrast space. 

\textbf{Exozodiacal disk sensitivity.} While we did not detect any extended emission, injection-recovery tests with model exozodi disks show that our observations are sensitive to extended emission coplanar with \acenAB, around \acenA, at an unprecedented level of \added{$\approx$5--8$\times$} the Solar System's zodiacal cloud. This corresponds to the deepest limits on exozodi achieved for any stellar system to date and is a factor of $\sim$5--10 improvement over current state-of-the-art limits from nulling interferometry. The tremendous gain is a result of JWST's ability to resolve \acenA's habitable zone (due to the star's proximity) and its excellent stability, which enables accurate subtraction of the stellar contribution.

\textbf{Detection of planet candidate \objS.} A comprehensive search for point sources in the vicinity of \acenA using both the classical median-RDI and joint PCA-KLIP subtraction strategies revealed one candidate, \objS, $\approx$1\farcs5 East of \acenA, 3.5 mJy in brightness, and detected at a S/N between 4 and 6 in August 2024. To investigate the nature of \objS, we performed several tests, exploring the following possibilities:

\begin{enumerate}
    \item \textbf{Is \objS a detector-level artifact in the \acenAB integrations?} The independent detection of \objS in multiple subsets (first half, second half, even frames, odd frames, individual integrations) of the \acenAB data and the absence of any jumps in mean pixel brightness in an aperture centered on \objS's location in the  individual \acenAB integrations shows that it is not likely a detector-level artifact (\S\ref{sec:subset}).

    \item \textbf{Is \objS a PSF subtraction artifact from \acenB?} \objS's signal persists both as the number of subsections in the annulus of joint PCA-KLIP reduction and the number of principal components used increases (\S\ref{sec:pcs-vary}). Importantly, \objS behaved distinctly than \emph{A}1, a point-like source identified to be a residual PSF subtraction artifact from \acenB, in the above tests. \emph{A}1 was not robust to changes in reduction parameters. Additionally, the \acenB PSF diffraction feature near the location of \objS in the August 2024 observation was also present in the February 2025 observation, but in the latter case, did not create a PSF subtraction artifact in the reductions (\S\ref{sec:feb_diff}). This is an independent argument that \objS is not likely a PSF subtraction artifact from \acenB.  

    \item \textbf{Is \objS an artifact from the \emus reference integrations?} The detection of \objS in all reductions performed by iteratively excluding frames from a single dither position (``leave-one-out" analysis, \S\ref{sec:leave-out}) shows that \objS is not likely an artifact from the on-axis \emus observations.

    \item \textbf{Is \objS a background or foreground object?} Taking advantage of \acenA's large proper motion ($\approx$3\farcs7 yr$^{-1}$), we show that no point source is recovered at the location expected for \objS as a stationary background object in both the February 2025 and April 2025 observations (\S\ref{sec:background-s1a}). The detection of \objS in the first and last integration of the $\approx2.5$ hour observing sequence, without any change in position, shows that \objS is not a high-proper motion foreground object (\S\ref{sec:subset}). \added{Additionally, no sources are present at the location of \acenA, during the August 2024 observations, according to the Minor Planet Catalog.} 

    \item \textbf{Is \objS re-detected as an orbiting planet in follow-up observations?} We did not recover \objS in a comprehensive search for point sources in the February 2025 and April 2025 observations. If \objS is \added{astrophysical in nature} (semimajor axis $\sim$1.5--2.5 au, orbital period $\sim$1.75--3.75 years; Paper I), it is expected to exhibit significant orbital motion between August 2024 and the above epochs. In this scenario, \objS has moved to region of poor sensitivity (e.g., within the coronagraph inner working angle or near the transition boundary), where it is undetectable. We generated 2D sensitivity maps for the February 2025 and April 2025 observations to enable constraints on the possible orbits of \objS consistent with a null detection in both epochs.
\end{enumerate}

Together, the above results present a case for \objS as a planet, as opposed to an image artifact or contaminating astrophysical source, from a data reduction standpoint. An independent astrophysical case can be made for the plausibility of \objS as a planet and is the topic of Paper~I (Beichman \& Sanghi et al. 2025, in press). Because the August epoch had only one successful observation at a single roll angle, it is not possible to confirm that \objS is a bona fide planet. Nevertheless, \objS merits careful consideration as a candidate signal. 

\added{If astrophysical in nature}, \objS's non-detection in two follow-up observations (February~2025 and April~2025), conducted within one year of candidate identification (August~2024), underscores the challenge of confirming short-period ($\sim$few years) planets that exhibit significant (but, a priori, unknown) orbital motion, on the timescales of standard observing cycles, with direct imaging. Indeed, considering \objS together with VLT/NEAR candidate $C1$, Paper I shows that $\sim$50\% of dynamically stable orbits, fit to the two astrometric points, would be consistent with non-detections in both follow-up observations. In such situations, several visits are needed to confirm planet candidates. Additional observations, with JWST or upcoming facilities \added{(e.g., the Roman Coronagraph Instrument, the European Extremely Large Telescope, the Giant Magellan Telescope, the TOLIMAN Space Telescope)}, are thus required to re-detect candidate \objS before making the dramatic and exciting claim that a giant planet orbits our nearest Sun-like star. If \objS is confirmed as a new ``$\alpha$~Cen~Ab", it would be the nearest (1.33 pc), oldest ($\sim$5 Gyr), coldest ($\approx$225~K, Paper~I), \added{shortest period ($\approx$~2--3~years, Paper~I)}, and lowest-mass ($\lesssim$~200~\mearth, Paper I) planet to be directly imaged around a solar-type star, to date. It would join a growing list planets, with temperatures more comparable to our own Jupiter, to be directly imaged and amenable to follow-up spectroscopic observations: $\epsilon$~Ind~Ab \citep{matthews_temperate_2024}, TWA~7b \added{\citep{lagrange_evidence_2025, crotts_follow-up_2025}}, 14~Her~c \citep{gagliuffi_jwst_2025}, \added{and the candidate/confirmed planets orbiting white dwarfs \citep{mullally_jwst_2024, limbach_miri_2024, limbach_thermal_2025}.}

\begin{acknowledgments}
The STScI support staff provided invaluable assistance in the planning and execution of this program. In particular, we thank George Chapman and the FGS team for their dedicated work in finding and vetting guide stars for this program and Wilson Joy Skipper and the short- and long-range planning teams for their contributions to this challenging observational program. The STScI's Director's office provided strong support for this program, from its initial selection as a high-risk, high-reward project, granting time to conduct test observations needed to validate the target acquisition strategy, to the execution of the follow-up DDT programs. \added{We thank the referee for a prompt report and helpful comments that improved this manuscript.}

This material is based upon work supported by the National Science Foundation Graduate Research Fellowship under Grant No.~2139433. Part of this work was carried out at the Jet Propulsion Laboratory, California Institute of Technology, under a contract with the National Aeronautics and Space Administration (80NM0018D0004). Program PID\#1618 is supported through contract JWST-GO-01618.001. NG and EC acknowledge funding by the European Union (ERC, ESCAPE, project No 101044152). Views and opinions expressed are however those of the author(s) only and do not necessarily reflect those of the European Union or the European Research Council Executive Agency. Neither the European Union nor the granting authority can be held responsible for them. All of the data presented in this paper were obtained from the Mikulski Archive for Space Telescopes (MAST) at the Space Telescope Science Institute. The specific observations analyzed can be accessed via
\dataset[doi:10.17909/v8nv-vx17]{http://dx.doi.org/10.17909/v8nv-vx17} for the August 2024 observations,\dataset[doi:10.17909/cb0x-rn85]{http://dx.doi.org/10.17909/cb0x-rn85} for the February 2025 observations, and\dataset[doi:10.17909/3z9q-9f65]{http://dx.doi.org/10.17909/3z9q-9f65} for the April 2025 observations. STScI is operated by the Association of Universities for Research in Astronomy, Inc., under NASA contract NAS5-26555. Support to MAST for these data is provided by the NASA Office of Space Science via grant NAG5-7584 and by other grants and contracts. This research has made use of NASA's Astrophysics Data System. Software citation information aggregated using \texttt{\href{https://www.tomwagg.com/software-citation-station/}{The Software Citation Station}} \citep{wagg_streamlining_2024, wagg_tomwaggsoftware-citation-station_2024}.
\end{acknowledgments}

\begin{contribution}
A.~Sanghi led the writing and submission of this manuscript under the guidance of C.~Beichman and D.~Mawet. W.~Balmer conducted the JWST pipeline processing of the MIRI observations. A.~Sanghi led the post-processing analysis of the MIRI observations for point sources and extended emission. N.~Godoy, M.~Sommer, M.~Wyatt, and E.~Choquet performed the S/N calculations for extended emission. L.~Pueyo, A.~Boccaletti, and J.~Llop-Sayson conducted independent post-processing of the MIRI observations to validate the results. K.~Wagner, A.~Bidot, and P.~O.~Lagage provided advice on data reduction. Dust emission models for the exozodiacal disk were developed by M.~Sommer and M.~Wyatt. P.~Kervella provided the ephemeris for \acenA. All authors discussed the results and commented on the manuscript.
\end{contribution}

\facility{JWST(MIRI)}

\software{\texttt{astropy
} \citep{astropy_collaboration_astropy_2013, astropy_collaboration_astropy_2018, astropy_collaboration_astropy_2022}, \texttt{matplotlib
} \citep{hunter_matplotlib_2007}, \texttt{numpy
} \citep{harris_array_2020}, \texttt{pandas
} \citep{mckinney_data_2010, team_pandas-devpandas_2025}, \texttt{python
} \citep{van_rossum_python_2009}, \texttt{scipy
} \citep{virtanen_scipy_2020, gommers_scipyscipy_2023}, \texttt{astroquery
} \citep{ginsburg_astroquery_2019, ginsburg_astropyastroquery_2024}, \texttt{scikit-image
} \citep{van_der_walt_scikit-image_2014}, \texttt{STPSF
} \citep{perrin_simulating_2012, perrin_updated_2014}, \texttt{jwst} \citep{bushouse_jwst_2025}, \texttt{pyKLIP} \citep{wang_pyklip_2015}, \texttt{vip} \citep{gomez_gonzalez_vip_2017, christiaens_vip_2023}, \texttt{spaceKLIP} \citep{kammerer_performance_2022, carter_jwst_2023, carter_spaceklip_2025}, and \texttt{webbpsf\_ext} \citep{leisenring_webbpsf_2025}.}
        
\appendix
\label{sec:app}
\restartappendixnumbering

\section{To Bin or Not to Bin: Example Comparison with February 2025 Observations}
\label{sec:app-bin}

\begin{figure*}[tb!]
\centering
\includegraphics[width=\textwidth]{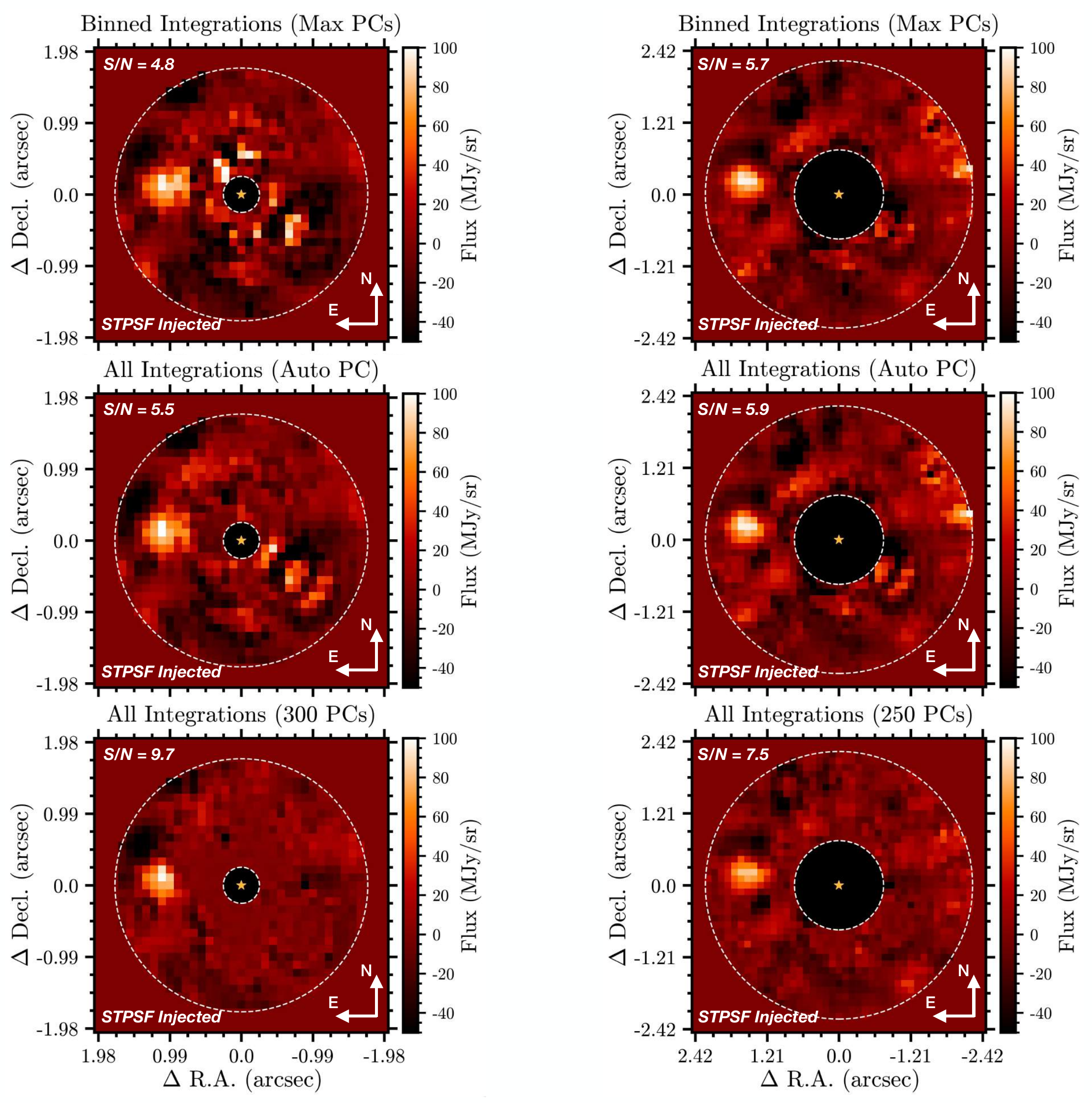}
\caption{\emph{Left column:} reduced images showing the recovery of a synthetic \webbpsf model injected at a separation of 1\arcsec\ and position angle of $80^\circ$ in the February 2025 observations. From top to bottom are reductions with a binned reference library, a reference library with all the integrations used and the number of PCs automatically selected, and a reference library with all the integrations used and the number of PCs selected to maximize the S/N of the injected companion. \emph{Right column:} same as the left column for injections at a separation of 1\farcs5. Using all the integrations results in a higher S/N recovery of the injected source compared to binning the integrations.}
\label{fig:bin_vs_nobin}
\end{figure*}

\added{For each observation epoch, we obtained 400 integrations per dither position in the \emus on-axis observations and 1250 integrations per \emus off-axis observation. An interesting question to investigate is whether we should median-combine (bin) all integrations of a given \emus (on-/off-axis) reference observation before PSF subtraction. The primary reason to retain all integrations as part of the reference library would be to leverage the frame-to-frame PSF diversity to improve PSF subtraction with PCA-KLIP. Here, we test the above through an example PSF injection-recovery test conducted with the February 2025 observations. We inject a 3 mJy \webbpsf model at separations of 1\arcsec\ and 1\farcs5 at a position angle of $80^\circ$ in the raw \acenAB cube. Three types of reductions are performed with PCA-KLIP applied to a 3 FWHM diameter annular region centered on the injected source's location: (1)~all the off-axis and each dither position of the on-axis \emus reference integrations are binned (mean-stacked) to a single frame (total number of reference frames after binning = 1 off-axis + 9 on-axis + 9 on-axis) and the number of principal components (PCs) is manually chosen to maximize the injected companion S/N; (2)~all \emus reference integrations are used without binning (total number of reference frames = 1250 off-axis + $9\times400$ on-axis + $9\times400$ on-axis) and the number of PCs is automatically chosen as described in \S\ref{sec:auto-pc}; and (3)~all \emus reference integrations are used without binning and the number of PCs is manually chosen to maximize the injected companion S/N. In the third case, we noticed that at very high PCs (approaching the maximum) artifacts are created at the pixel level which artificially inflate the estimated S/N, even though the reduced images show that the synthetic companion has been severely over-subtracted. Thus, the reduced images had to be visually inspected to reliably select the optimal number of PCs. This issue makes it impractical to manually tune the number of PCs in our full set of injection-recovery tests to generate 2D sensitivity maps (\S\ref{sec:sensitivity}).} 

\added{For both injection separations, we find that using all the integrations without binning recovers the synthetic companion at a higher S/N as compared to the reductions where the reference integrations were binned (Figure~\ref{fig:bin_vs_nobin}; 19 PCs, the maximum possible, was preferred for the binned reductions). Further, we find that manually tuning the number of PCs can further improve confidence in the detectability of a point source, compared to auto-PC selection (Figure~\ref{fig:bin_vs_nobin}), as was also seen in the analysis presented in \S\ref{sec:pcs-vary} for the August 2024 observations. While the results of the comparison are only presented for two injection locations (to serve as examples), we find that the same conclusions generally hold true at other separations and position angles of injection as well. Given these results, we chose not to bin the individual reference integrations before PSF subtraction.}

\added{The temporal behavior of wavefront variations of JWST have been reported in detail by \citet{telfer_empirical_2024}. At the timescale of our single integrations, $\sim$7.5~seconds, shorter vibrations dominate at the $\sim$200~pm level. At the timescale of minutes, the ISIM Electronics Compartment (IEC) oscillations dominate with wavefront variations of the order of $\sim$1~nm. The effect of the IEC oscillations can be seen in the power spectrum of the principal components computed from the set of science or reference integrations. This explains why using a single integration PSF library performs better. The above analysis will be discussed in more detail in future work.}

\section{Recovery of Known Background Sources}
\label{sec:recover-back}

\begin{figure}[ht!]
\centering
\includegraphics[width=\textwidth]{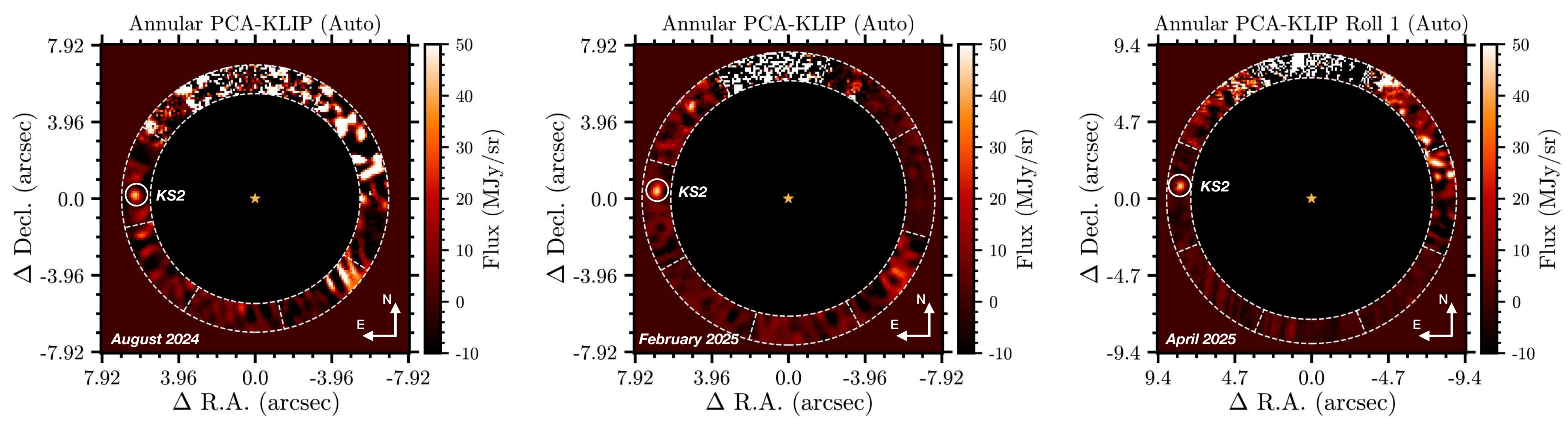}
\caption{From left to right, joint PCA-KLIP reduced images for the August 2024, February 2025, and April 2025 (roll 1) epoch observations showing the detection of the $\sim$1 mJy background source KS2 (circled) from \citet{kervella_close_2016}.}
\label{fig:bckg-aug}
\end{figure}

\citet{kervella_close_2016} presented a catalog of the expected close conjunctions between the \acenAB system and stars in the background. Among them, sources S2 and S5 (hereafter referred to as KS2 and KS5) are expected to be in the vicinity of \acenA during the time of our observations. Indeed, KS5 is seen in the \acenAB integration even before PSF subtraction at the edge of the detector (Figure \ref{fig:fullfr}). This is the bright source that \acenA is rapidly moving towards in upcoming years with a distance of closest approach $\rho_{\rm min} = 0\farcs015 \pm 0\farcs135$ in $\sim$2028 \citep{kervella_close_2016}. KS2 is expected to be located $\approx$6\farcs3, $\approx$6\farcs9, and $\approx$8\farcs3 East from \acenA for the August 2024, February 2025, and April 2025 observing dates, respectively, and has a low F1550C flux of $\sim$1~mJy. We use this source as a test object to check if PCA-KLIP can recover it. We performed joint PCA-KLIP in an annulus of width 3 FWHM divided into eight subsections, centered at the radial separation expected for KS2 in each epoch, and with the number of PCs automatically selected as discussed in \S\ref{sec:auto-pc}. KS2 is clearly detected at the expected location in each epoch (Figure \ref{fig:bckg-aug}). For the April dataset specifically, KS2 is only detected in the Roll 1 observation as it lies along the 4QPM transition boundary for the second roll.

\bibliography{\string references.bib}
\bibliographystyle{aasjournalv7}
\end{document}